\documentclass[prb,reprint,amsmath,amssymb,floatfix,superscriptaddress]{revtex4-1}
\usepackage{graphicx}
\usepackage{relsize}
\usepackage{hyperref}  
\usepackage{enumerate}
\usepackage{bbold}
\usepackage{xcolor}
\renewcommand{\vec}[1]{\mathbf{#1}}

\begin{document}

\title{Effects of deep superconducting gap minima and disorder on residual thermal transport in $\mathrm{Sr_2 Ru O_4}$}

\author{John F. Dodaro}
\affiliation{Department of Physics, Stanford University, Stanford, CA 94305-4060, USA}
\author{Zhiqiang Wang}
\affiliation{Department of Physics and Astronomy, McMaster University, Hamilton, Ontario, L8S 4M1, Canada}
\author{Catherine Kallin}
\affiliation{Department of Physics and Astronomy, McMaster University, Hamilton, Ontario, L8S 4M1, Canada}
\affiliation{Canadian Institute for Advanced Research, Toronto, Ontario M5G 1Z8, Canada}
\date{\today}

\begin{abstract}
Recent thermal conductivity measurements on $\mathrm{Sr_2 Ru O_4}$ [E. Hassinger \textit{et al.}, Phys. Rev. X \textbf{7}, 011032 (2017)] were interpreted as favoring
a pairing gap function with vertical line nodes while conflicting with chiral $p$-wave pairing.  Motivated by
this work we study the effects of deep superconducting gap minima on impurity induced quasiparticle thermal transport in chiral $p$-wave models of $\mathrm{Sr_2 Ru O_4}$.
Combining a
self-consistent T-matrix analysis and self-consistent Bogoliubov-de-Gennes calculations, we show that the dependence of the residual
thermal conductivity on the normal state impurity scattering rate can be quite similar to the $d$-wave pairing state that was shown to fit the thermal conductivity measurements, provided the normal state impurity scattering rate
is large compared with the deep gap minima. Consequently, thermal conductivity measurements on $\mathrm{Sr_2RuO_4}$ can be reconciled with a chiral
$p$-wave pairing state with deep gap minima. However, the data impose serious constraints on such models and these constraints are examined in the context of several different
chiral $p$-wave models.
\end{abstract}


\maketitle

\section{Introduction}
Understanding unconventional superconductors, including both their pairing symmetries and mechanisms, has been a great challenge.  Among the many unconventional
superconductors discovered, $\mathrm{Sr_2 Ru O_4}$ is thought to realize a spin triplet chiral $p$-wave superconductor~\cite{Mackenzie2003,Kallin2009,Kallin2012,Kallin2016,Maeno2012}.
However, despite more than twenty years of study the exact nature of its superconducting order parameter remains a puzzle, which is partially due to conflicting interpretations
of different experiments~\cite{Mackenzie2017}. The proposal of spin triplet chiral $p$-wave pairing has been supported by many experiments~\cite{Maeno2012}. Spin susceptibility measurements on
$\mathrm{Sr_2 Ru O_4}$, including nuclear magnetic resonance~\cite{Ishida1998,Ishida2015} and polarized neutron scattering~\cite{Duffy2000}, do not see any drop of the
electronic spin susceptibility below the superconducting transition temperature $T_c$, consistent with spin triplet pairing.  Further support for chiral $p$-wave pairing
comes from the spontaneous time reversal symmetry breaking revealed by muon spin relaxation~\cite{Luke1998,Luke2000} and polar Kerr effect
measurements~\cite{Xia2006}. In its simplest form, chiral $p$-wave pairing gives rise to a full gap on the entire Fermi surface (FS) sheets of $\mathrm{Sr_2 Ru O_4}$,
which, however, is incompatible with experiments probing the low energy excitations. These include specific heat~\cite{Nishizaki1999,Nishizaki2000,Deguchi2004,Deguchi2004a},
ultrasound~\cite{Lupien2001} and penetration depth measurements~\cite{Bonalde2000}, all of which imply that low energy excitations exist deep inside the superconducting
state.  In order to account for these experiments, chiral $p$-wave pairing gap functions with deep minima (or accidental nodes) have been
proposed~\cite{Miyake1999} and supported by microscopic calculations~\cite{Nomura2005,Raghu2010,Wang2013,Scaffidi2014}.

Since deep gap minima are not protected by any symmetry, unlike symmetry-enforced nodes, their occurrence in $\mathrm{Sr_2RuO_4}$ requires some explanation~\cite{Raghu2010,Firmo2013}. In the calculations of Refs.~\onlinecite{Raghu2010,Scaffidi2014} deep gap
minima appear on the $\alpha$ and $\beta$ FS sheets, which are generated by small hybridizations of the $\mathrm{Ru}$ $4d_{xz}$ and $4d_{yz}$ quasi-one-dimensional ($1d$) bands.
These gap minima are vertical as they exist at all values of $k_z$. If we ignore the
hybridization as well as small couplings to the $\gamma$-band, the near nesting of the quasi-$1d$ bands and the corresponding peak in the anti-ferromagnetic
spin fluctuation spectrum at the nesting wavevectors~\cite{Sidis1999} favor $p$-wave superconducting order parameters with accidental nodes on each quasi-$1d$ band~\cite{Raghu2010,
Firmo2013,Scaffidi2015}.
When the small hybridization is included, those accidental nodes are transformed into small gaps with magnitude of order
$\Delta_{\mathrm{1d}}(t^{\prime\prime}/t)^2\sim 0.01 \Delta_{\mathrm{1d}}$~\cite{Raghu2010,Firmo2013}. Here $\Delta_{\mathrm{1d}}$ is the gap magnitude on the
quasi-$1d$ bands in the absence of hybridization, $t^{\prime\prime}$
is the next-nearest neighbor inter-orbital hopping that mixes the two quasi-$1d$ bands,\footnote{The mixing between the two quasi-$1d$ bands actually arises from a combination of inter-orbital hopping $t^{\prime\prime}$ and 
spin-orbit coupling (SOC), $\lambda_{\mathrm{SOC}}$. The $\lambda_{\mathrm{SOC}}$ extracted from LDA band splitting at the $2d$ Brillouin zone center
is not small compared with the hopping parameter $t$~\cite{Haverkort2008,Liu2008}; however, its effect on the FS
is perturbatively small near the $(1,1)$ direction~\cite{Haverkort2008}, which suggests a smaller $\lambda_{\mathrm{SOC}}$ to be used in a tight-binding calculation
to model the physics near the FS, as in Ref.~\onlinecite{Scaffidi2014}\,.} and $t$ is the nearest neighbor hopping that gives rise to the quasi-$1d$ bands.
Therefore the accidental nodes become deep gap minima or ``near-nodes" with hybridization.
In other words, an isotropic chiral $p$-wave is not expected in a lattice calculation.
We emphasize that the occurrence of deep gap minima on the $\alpha$ and $\beta$
bands results from the quasi-$1d$ nature of their band structures rather than a fine tuning of the underlying microscopic interactions~\cite{Raghu2010,Firmo2013}.

However, while the substantial low energy density of states arising in models of chiral $p$-wave with near-nodes can explain specific heat data on $\mathrm{Sr_2RuO_4}$~\cite{Firmo2013}, such models have been challenged by thermal conductivity measurements~\cite{Hassinger2017}.
In Ref.~\onlinecite{Hassinger2017}, the dependence of
the residual thermal conductivity on the normal state impurity scattering rate has been shown to follow the $d$-wave pairing prediction~\cite{Durst2000,Sun1995,Graf1996}
(with vertical line nodes) within experimental error bars.
In particular, the available in-plane residual thermal conductivity data obtained from different samples with different amount of disorder
(see Fig.1 of Ref.~\onlinecite{Hassinger2017} and Fig.2 of Ref.~\onlinecite{Suzuki2002}) suggests that the residual thermal conductivity extrapolates to a large nonzero
constant as the impurity scattering rate decreases to zero.  This is consistent with the well-known universal thermal transport~\cite{Durst2000,Sun1995,Graf1996} of
a superconducting state with line nodes; while it is completely different from what is expected for the isotropic chiral $p$-wave case.
For isotropic chiral $p$-wave, the residual thermal conductivity becomes vanishingly small~\cite{Maki1999,Maki2000} in the zero impurity scattering limit,
since the number of low energy excitations available for heat transport decreases rapidly below the isotropic superconducting gap and impurity induced sub-gap states
are localized~\cite{Balatsky2006}.

Despite their consistency with the thermal conductivity data~\cite{Hassinger2017}, vertical line nodes are, generically, not compatible with a time reversal symmetry breaking superconducting state in $\mathrm{Sr_2RuO_4}$ due to Blount's theorem~\cite{Blount1985}.
Therefore, it is useful to consider vertical near-nodes in order to reconcile the time reversal symmetry breaking with the thermal transport measurements.

Although it is well known that near-nodes or accidental nodes in an $s$-wave superconductor can be easily washed out by impurity scattering~\cite{Borkowski1994}, the effect of disorder on accidental nodes or near-nodes in a non-$s$-wave superconductor
has not received much attention, largely because non-$s$-wave superconductors typically have symmetry protected nodes that dominate the low-temperature behavior, while accidental
nodes or near-nodes are much less common and typically require fine-tuning of microscopic parameters. However, this issue becomes important in a multi-component non-$s$-wave superconductor
like chiral $p$-wave in $\mathrm{Sr_2RuO_4}$ that does not have any symmetry protected nodes.

A key point is that $s$-wave and non-$s$-wave behave very differently in this respect.
Unlike in an $s$-wave superconductor, accidental nodes or deep minima in a non-$s$-wave superconductor can be robust to impurity scattering
and lead to a dependence of the residual thermal conductivity on the normal state impurity scattering rate $\Gamma_N$, similar to that of the $d$-wave pairing
case, provided $\Gamma_N\gtrsim \Delta_{\min}$ (we set $\hbar=1$). Here $\Delta_{\min}$ is the minimum zero temperature gap magnitude. The explanation for the difference is essentially the same as the explanation for
why $s$-wave is robust to non-magnetic potential scattering disorder, while non-$s$-wave is easily destroyed by such disorder. At low temperature, impurities scatter Bogoliubov quasiparticles
around the Fermi surface, effectively averaging the gap function (not the gap magnitude) around the Fermi surface, which leads to a robust, more isotropic gap for
$s$-wave and to a reduced gap at all $\vec{k}$ for non-$s$-wave pairing. In the self-consistent T-matrix formalism, used in this paper, this impurity-averaging
effect adds a self-energy off-diagonal in the Nambu particle-hole space, $\Sigma_{\mathrm{o.d.}}(\omega)$, to the original anisotropic clean system gap function
$\Delta(\vec{k},\omega)$: $\Delta_{\mathrm{eff}}(\vec{k})=\Delta(\vec{k})+\Sigma_{\mathrm{o.d.}}(\omega)$.
To first order in the impurity concentration, $\Sigma_{\mathrm{o.d.}}(\omega) \propto \langle \mathcal{F}(\omega,\vec{k}) \rangle_{\mathrm{FS}}$, where $\mathcal{F}$
is the clean system anomalous Green's function and $\langle \cdots \rangle_{\mathrm{FS}}$ denotes an average over the
Fermi surfaces. For $s$-wave, $\Sigma_{\mathrm{o.d.}}\ne 0$ and effectively gaps out the accidental nodes.
For non-$s$-wave, the FS average is zero, so $\Sigma_{\mathrm{o.d.}}=0$ and the disorder averaged gap function (while reduced overall) has the same anisotropy
and deep gap minima as the clean $\Delta(\vec{k})$. Furthermore, if $\Gamma_N \gtrsim \Delta_{\min}$, the impurity induced states below $\Delta_{\min}$ are delocalized.
Consequently, the effect of disorder on near-nodes or accidental nodes in a non-$s$-wave superconductor is similar
to the effect on symmetry protected nodes provided $\Gamma_N \gtrsim \Delta_{\min}$, which only requires a tiny amount of disorder if $\Delta_{\min}$ is very small.

In this paper, we support the above arguments with explicit residual thermal conductivity calculations for different chiral $p$-wave pairing models,
proposed for $\mathrm{Sr_2 Ru O_4}$, that
have deep minima and study in some detail the constraints that experiments place on such models.  Our calculations use the self-consistent T-matrix approximation (SCTA)
and the self-consistent Bogoliubov de Gennes (BdG) equations.
An analysis of the residual thermal conductivity within the SCTA
in Appendix~\ref{app:localization_kappa} shows that the substantial residual thermal conductivity at $\Gamma_N\gtrsim \Delta_{\min}$ also implies delocalized zero-energy Bogoliubov
quasiparticle states; while for $\Gamma_N\lesssim \Delta_{\min}$ the zero energy states tend to localize.
Since SCTA is only approximate, we also analyze the effects of disorder using self-consistent BdG which includes scattering effects beyond the SCTA and allows local order
parameter variations. These calculations confirm which states are localized or delocalized and show that our conclusions remain valid beyond the SCTA.

The effects of impurity scattering on chiral $p$-wave pairing with deep minima have been studied for $\mathrm{Sr_2 Ru O_4}$ previously in
Refs.~\onlinecite{Miyake1999,Nomura2005}.  However, Ref.~\onlinecite{Miyake1999} focuses on the impurity induced residual density of states and its
thermodynamic signatures rather than transport; Ref.~\onlinecite{Nomura2005} has calculated thermal conductivity in the presence of disorder within the SCTA, but
only for a particular impurity concentration. Neither studies the effect of different amount of disorder on the residual thermal conductivity which is the focus of this paper.
Furthermore, the impurity concentration considered in Ref.~\onlinecite{Nomura2005} is too small for a direct comparison to experiments~\cite{Suzuki2002,Hassinger2017}
(for more detailed discussions, see Sec.~\ref{sec:SCTA_2band}).

Although, deep gap minima in chiral $p$-wave can lead to a residual thermal conductivity similar to $d$-wave, the fact that the experimental data is well fit by assuming
$d$-wave on all three bands does place considerable constraints on models of chiral $p$-wave with near-nodes.
These constraints are explored here by considering several different chiral $p$-wave models, including the possibility of
horizontal line nodes which have been invoked to explain some experiments on $\mathrm{Sr_2 Ru O_4}$~\cite{Hasegawa2000,Zhitomirsky2001,Annett2002,Wysokinski2003,Litak2004,Koikegami2003,Kittaka2018}.

The paper is organized as follows.  In Sec.~\ref{sec:SCTA} we describe the residual thermal conductivity calculation for various pairing models~\cite{Raghu2010,Scaffidi2014,Scaffidi2015,Wysokinski2003}
and compare the results with experiments~\cite{Suzuki2002,Hassinger2017} in Sec.~\ref{sec:SCTA_results}.  In Sec.~\ref{sec:BdG} we present a self-consistent BdG analysis
which confirms the SCTA and shows that the impurity induced states below $\Delta_{\min}$ are
delocalized for $\Gamma_N \gtrsim \Delta_{\mathrm{min}}$.  Sec.~\ref{sec:conclusion} contains conclusions and further discussions.
Appendix~\ref{app:gapprofile} and~\ref{app:kappa} provide some technical computational details and further discussions of the various models used
in our calculations. Appendix~\ref{app:localization_kappa} contains a discussion of localization effects on the residual thermal conductivity within the SCTA.
Although the main body of the paper is focused on thermal conductivity, in Appendix~\ref{app:DOS}, we also contrast the effect of disorder on the low energy
density of states of a non-$s$-wave superconductor with near-nodes to that of an $s$-wave supercondutor, employing self-consistent BdG calculations.

\section{Residual thermal conductivity in the SCTA} \label{sec:SCTA}
We first outline the general procedure of the residual thermal conductivity calculation within the SCTA for a general BdG Hamiltonian,
$\hat{H}_{\mathrm{BdG}}$, which may consist of two or three orbitals/bands.

Consider an $N$-band BdG Hamiltonian, $\hat{H}_{\mathrm{BdG}}(\vec{k})$, which is a $2N\times 2N$ matrix. We denote all matrix quantities with a hat.
The clean
system Green's function, $\hat{G}_0(i\omega_n,\vec{k})$, is defined from its inverse:
\begin{gather}
\hat{G}_0^{-1}(i\omega_n,\vec{k}) \equiv i \omega_n - \hat{H}_{\mathrm{BdG}}(\vec{k}), \label{eq:G0inv}
\end{gather}
where $\omega_n=(2n+1)\pi T$ is the fermionic Matsubara frequency and $T$ the temperature which will be set to zero at the end. The effect of impurity scattering
on the Bogoliubov quasiparticles is included via an impurity self energy, $\hat{\Sigma}(i\omega_n,\vec{k})$. The momentum $\vec{k}$ is still a good quantum number
because the translational symmetry is restored after the impurity potential configuration average. We take the impurity scattering potential to be isotropic and completely $\vec{k}$ independent,
so, in $\vec{k}$ space, $\hat{V}_{\mathrm{imp}}=V_0  \mathbb{1}_{N\times N} $, where $V_0$ is a constant and $\mathbb{1}_{N\times N}$
is the identity matrix in the orbital sub-space. As argued in Refs.~\onlinecite{Mackenzie1996,Mackenzie1998,Miyake1999}, the impurity scattering in $\mathrm{Sr_2 Ru O_4}$ is in the unitary scattering limit:
$V_0\rightarrow \infty$, which will be taken in our calculation.
Since $\hat{V}_{\mathrm{imp}}$ does not depend on $\vec{k}$, $\hat{\Sigma}(i\omega_n,\vec{k})\equiv\hat{\Sigma}(i\omega_n)$ is independent of $\vec{k}$ as well, and
within the SCTA, is given by~\cite{Hirschfeld1988,Borkowski1994}
\begin{gather}
\hat{\Sigma}(i\omega_n)=\frac{n_i \; \hat{V}_{\mathrm{imp}} \; \tau_3}{1-\overline{\hat{G}}(i \omega_n) \hat{V}_{\mathrm{imp}} \; \tau_3}, \label{eq:Tmatrix}
\end{gather}
where $n_i$ is the impurity concentration and
 $\tau_3$ is the $z$-component Pauli matrix of the particle-hole Nambu sub-space.
For a general superconductor, all $2N\times 2N$ matrix elements of  $\hat{\Sigma}$ can be nonzero.
However, for a non-s wave superconductor, the anomalous part of $\hat{\Sigma}$ is always identically zero~\cite{Hirschfeld1988,Borkowski1994},
and $\hat{\Sigma}$ has at most $2 N^2$ nonzero elements. For $V_0\rightarrow \infty$, the impurity scattering strength
 is solely characterized by $n_i$, which is directly proportional to the normal state impurity scattering rate, $\Gamma_N$.
 In the denominator of Eq.~\eqref{eq:Tmatrix}, $\overline{\hat{G}}(i\omega_n)$
 is the $\vec{k}$-space averaged Green's function
\begin{gather}
\overline{\hat{G}}(i\omega_n)=\frac{1}{N_{\vec{k}}}\sum_{\vec{k}} \hat{G}(i\omega_n,\vec{k}), \label{eq:Gbar}
\end{gather}
where $\frac{1}{N_{\vec{k}}}\sum_{\vec{k}}$ means averaging over the first Brillouin zone and
$\hat{G}(i\omega_n,\vec{k})$ is the full disorder averaged Green's function, defined by
\begin{gather}
\hat{G}^{-1}(i\omega_n,\vec{k}) =\hat{G}_0^{-1}(i\omega_n,\vec{k}) -\hat{\Sigma}(i\omega_n). \label{eq:Ginv}
\end{gather}
For a given BdG Hamiltonian and $n_i$, Eqs.~\eqref{eq:G0inv}-\eqref{eq:Ginv} form a set of closed self-consistent
equations for the impurity self energy matrix $\hat{\Sigma}(i\omega_n)$ and can be solved numerically by iteration.

However, the non-magnetic impurity scattering is also pair breaking for non-s wave superconductors, and degrades the superconducting order parameter $\hat{\Delta}(\vec{k})$
that enters into the BdG Hamiltonian of the above equations. This is taken into account by supplementing
Eqs.~\eqref{eq:G0inv}-\eqref{eq:Ginv} with the superconducting gap equation for $\hat{\Delta}(\vec{k})$. We start with the gap function in the orbital basis, $\hat{\Delta}(\vec{k})$,
which is a diagonal matrix in all the models that we study: $\hat{\Delta}(\vec{k})_{\alpha,\beta}=\Delta_{\alpha}(\vec{k}) \delta_{\alpha,\beta}$ with orbital labels $\alpha=1,...,N$.
Then we perform a unitary transformation on $\hat{\Delta}(\vec{k})$ to obtain the gap functions in the band basis, $\hat{\Delta}^b(\vec{k})=\hat{U}^\dagger_{\vec{k}} \hat{\Delta}(\vec{k}) \hat{U}_{-\vec{k}}^*$,
where $\hat{U}_{\vec{k}}$ is the unitary matrix that diagonalizes the normal state Hamiltonian at the wavevector $\vec{k}$.
In general, $\hat{\Delta}^b(\vec{k})$ is not diagonal in the band basis, meaning some inter-band pairing has been included.
However, these inter-band pairing terms are small over most of the FS and, also, the lowest lying Bogoliubov quasiparticle energies do not
depend on them to leading order in the overall gap magnitude.\footnote{For the 2-band model in Sec.~\ref {sec:SCTA_2band} the
Bogoliubov quasiparticle energies were given in Ref.~\onlinecite{Taylor2013} in terms of orbital pairing gap functions, and can be
transformed to the band basis. Along the $\beta $ band FS, where the lowest lying quasiparticle energy is realized, one finds in weak
coupling, $E(\vec{k})^2=|\Delta _{\beta ,\mathrm{intra}}(\vec{k})|^2+\mathcal{O}(|\Delta|^4/E_{\alpha ,N}^2)$, where
$\Delta_{\beta ,\mathrm{intra}}(\vec{k})$ is the $\beta$ intra-band pairing gap function, and $E_{\alpha ,N}(\vec{k})$ is the $\alpha$-band
normal state energy dispersion. Note, along the $\beta$-band FS, $E_{\alpha ,N}(\vec{k})\ne 0$ and is on the order of the hopping parameters 
$t$, $t^{\perp}$, or $t^{\prime \prime }$, which are $\gg |\Delta |$.}
As a consequence, the inter-band pairing terms do not have any noticeable effect on the residual thermal 
conductivity.
Therefore, we will neglect them in the SCTA calculation so that there is only one pairing gap equation for each diagonal component of $\hat{\Delta}^b(\vec{k})$.
(However, we note that the BdG analysis in Sec.~\ref{sec:BdG} does include inter-band pairing.)
If we write these diagonal components as $\hat{\Delta}^b_{\alpha,\alpha}(\vec{k})=\Delta_\alpha f_{\alpha}(\vec{k})$, where $\Delta_{\alpha}$ is
the overall pairing magnitude
of the $\alpha$-th band and $f_{\alpha}(\vec{k})$ is the corresponding dimensionless gap function,
then the gap equations to be solved are given by
\begin{gather}
\Delta_{\alpha}^b = V_{\alpha} \; \pi T \sum_n \frac{1}{N_{\vec{k}}}\sum_{\vec{k}} f_{\alpha}^*(\vec{k}) \; \mathcal{F}_{\alpha}(i\omega_n,\vec{k}), \label{eq:BCS}
\end{gather}
where $\mathcal{F}_{\alpha}(i\omega_n,\vec{k})\equiv \hat{G}_{\alpha,\alpha+N}^b(i\omega_n,\vec{k})$ is the $\alpha$-th band anomalous Green's function.
The superscript $b$ in $\hat{G}^b_{\alpha,\alpha+N}$ means $\hat{G}^b$ is the band basis Green's function (disorder averaged), obtained from the orbital one
by $\hat{G}^b(i\omega_n,\vec{k})=\widetilde{U}_{\vec{k}}^\dagger \hat{G}(i\omega_n,\vec{k}) \widetilde{U}_{\vec{k}}$ with
 $\widetilde{U}_{\vec{k}}\equiv \mathrm{diag}\{\hat{U}_{\vec{k}},\hat{U}_{-\vec{k}}^*\}$.  $V_{\alpha}<0$ is the attractive pairing interaction strength for the $\alpha$-th band.
In writing down the above gap equation we have assumed that the effective pairing interaction for the $\alpha$-th band takes the factorizable form,
$V_{\mathrm{BCS}}^{\alpha}(\vec{q},\vec{k})=V_{\alpha} f_{\alpha}(\vec{q})
f^{*}_{\alpha}(\vec{k})$, such that it reproduces the desired pairing channel for the $\alpha$-th band. The magnitude of $V_{\alpha}$ is determined by the clean system pairing gap magnitude. Furthermore, we have assumed that the pairing interaction
is not affected by the impurity scattering.  Since the pairing magnitude,
$\Delta^b_{\alpha}$, is non-degenerate for different bands, in general,  we need to solve all $N$ pairing gap equations simultaneously. Also the critical impurity concentration,
$n_{i,c}$, is defined as the one at which all $\Delta^b_{\alpha}$ vanish.
Solving the coupled Eqs.~\eqref{eq:G0inv}-\eqref{eq:BCS} numerically by iteration (for $T=0$) we obtain both the disorder averaged Green's
functions and  the disorder averaged pairing gap magnitude.

With the above information we can compute the residual thermal conductivity $\kappa_0(n_i)$, defined by $
\frac{\kappa_0(n_i)}{T} \equiv \lim_{T\rightarrow 0,\Omega \rightarrow 0} \frac{\kappa(\Omega,T)}{T},
$
where $\kappa(\Omega,T)$ is the frequency $\Omega$ and temperature $T$ dependent thermal conductivity. Note that $\kappa_0$ depends on $n_i$.
Following Ref.~\onlinecite{Durst2000}, we start with the thermal current operator
matrix, $\hat{\vec{j}}^Q(\Omega,\vec{q}\approx 0)=\sum_{\vec{k},\omega}(\omega+\frac{\Omega}{2})\Psi^\dagger(\omega+\Omega,\vec{k}) \,
\nabla_{\vec{k}} \hat{H}_{\mathrm{BdG}}^{\mathrm{diag}} \,
\Psi(\omega,\vec{k})$, which depends on frequency, $\Omega$, and momentum, $\vec{q}$, but here the long wavelength limit is taken: $\vec{q}\rightarrow 0$.
The superscript ``diag" in $\hat{H}_{\mathrm{BdG}}^{\mathrm{diag}}$ means that the superconducting order parameter contribution~\cite{Durst2000} to the
thermal current velocity operator has been dropped, which is a very good approximation for $\mathrm{Sr_2RuO_4}$ since its superconducting gap is much smaller than the normal state band
parameters.
Then $\kappa_0/T$ can be computed from a thermal Kubo formula~\cite{Durst2000}.
The final result is
\begin{widetext}
\begin{gather}
\frac{\kappa_0(n_i)/T}{\pi^2 k_B^2 /3} = \frac{\pi}{2} \sum_{\vec{k}} \;
\mathrm{Tr}\bigg\{ \partial_{k_x} \hat{H}_{\mathrm{BdG}}^{\mathrm{diag}}(\vec{k}) \; \hat{\mathcal{A}}(0,\vec{k}) \;
\partial_{k_x} \hat{H}_{\mathrm{BdG}}^{\mathrm{diag}}(\vec{k}) \; \hat{\mathcal{A}}(0,\vec{k}) \bigg\}, \label{eq:kappa0}
\end{gather}
\end{widetext}
where $ \hat{\mathcal{A}}(\omega,\vec{k}) \equiv  \big\{ \hat{G}_{\mathrm{ret}}(\omega,\vec{k})-\hat{G}_\mathrm{adv}(\omega,\vec{k})\big\}/(-i\,2\pi) $
with the retarded/advanced Green's function given by
$\hat{G}_{\mathrm{ret/adv}}(\omega,\vec{k})=\hat{G}(i\omega_n \rightarrow \omega \pm i\delta,\vec{k})$.
We normalize $\kappa_0(n_i)/T$ by its value at the critical impurity concentration, $n_{i,c}$ (the corresponding normal impurity scattering rate is denoted as $\Gamma_c$).
Since $n_i \propto \Gamma_N$, then $\Gamma_N/\Gamma_c=n_i/n_{i,c}$ and also $\kappa_0(n_i)/\kappa_0(n_{i,c})=\kappa_0(\Gamma_N)/\kappa_0(\Gamma_c)$.

\section{SCTA results for different pairing models with deep minima} \label{sec:SCTA_results}
\subsection{$2$-band model} \label{sec:SCTA_2band}
\begin{figure}
\centering
\includegraphics[width=1.05\linewidth]{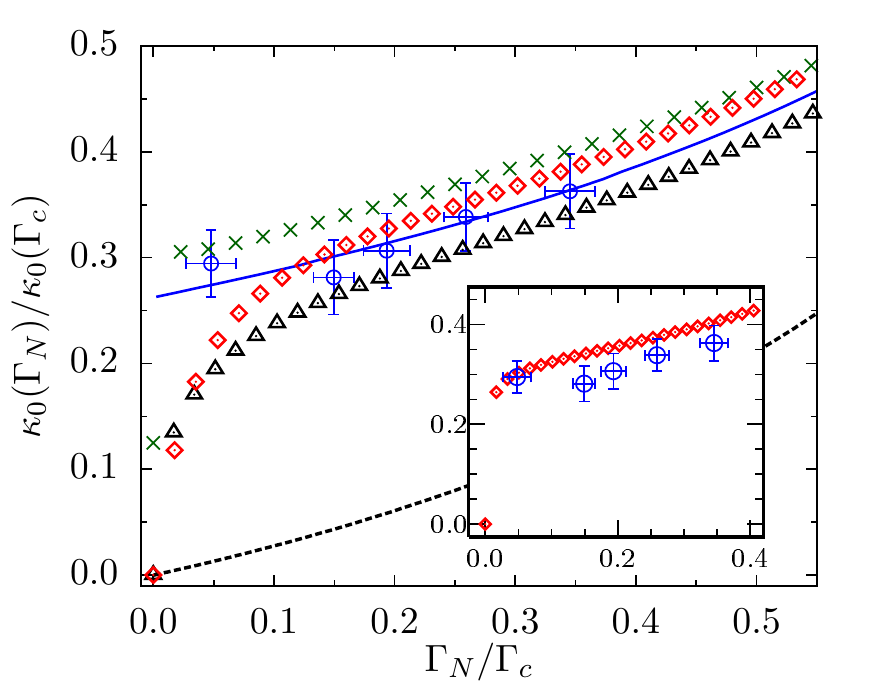}
\caption{Comparison of the residual thermal conductivity $\kappa_0(\Gamma_N)/\kappa_0(\Gamma_c)$ calculated for different theoretical models with the experimental
data~\cite{Suzuki2002, Hassinger2017}. Blue $\bigcirc$ with error
bars are the experimental data reproduced from Ref.~\onlinecite{Suzuki2002,Hassinger2017}; black dashed line is for the single-band isotropic $p_x+i p_y$ pairing obtained
for an isotropic FS~\cite{Maki1999,Maki2000};
blue solid line is for the single-band $d_{x^2-y^2}$ pairing with an isotropic FS~\cite{Sun1995,Sun1995a}; red $\diamondsuit$ are results for the $2$-band chiral $p$-wave pairing model with near-nodes from
Ref.~\onlinecite{Raghu2010}; dark green $\times$ are results for the
$3d$ chiral $p$-wave pairing model with both vertical near-nodes and horizontal line nodes; and the black $\bigtriangleup$ are results for the $3$-band model
with deep minima on both the $\beta$ and $\gamma$ bands (see Sec.~\ref{sec:SCTA_gamma}).
The inset compares the data of Refs.~\onlinecite{Suzuki2002,Hassinger2017}
with the $2$-band chiral $p$-wave model with a smaller orbital hybridization parameter $t^{\prime\prime}=0.05\, t$ (see text) from Ref.~\onlinecite{Rozbicki2011}.
}
\label{fig:kappaT}
\end{figure}

We consider several different chiral $p$-wave pairing models with deep gap minima that are relevant to $\mathrm{Sr_2RuO_4}$, providing details on each model as well as the corresponding
numerical results for the residual thermal conductivity. The first one is a simplified model,
the $2d$ $2$-band chiral $p$-wave pairing model proposed in Ref.~\onlinecite{Raghu2010}. In this model only the two quasi-$1d$ $d_{xz}$ and $d_{yz}$ orbitals
are considered. The BdG Hamiltonian is
\begin{gather}
\hat{H}_{\mathrm{BdG}} =
 \begin{pmatrix}
  \epsilon_a(\vec{k}) & g(\vec{k}) & \Delta_a(\vec{k}) & 0 \\
  g(\vec{k}) &  \epsilon_b(\vec{k}) & 0  &  \Delta_b(\vec{k}) \\
  \Delta_a^*(\vec{k}) & 0  &  -\epsilon_a(\vec{k}) & -g(\vec{k})   \\
 0  &   \Delta_b^*(\vec{k}) & -g(\vec{k}) &  -\epsilon_b(\vec{k})
 \end{pmatrix}, \label{eq:H2band}
\end{gather}
where $a,b$ stand for $d_{xz}$ and $d_{yz}$ orbitals, respectively. $\epsilon_a(\vec{k}) = - 2 t \cos k_x - 2 t^\perp \cos k_y -\mu$, $\epsilon_b(\vec{k}) = - 2 t \cos k_y - 2 t^\perp \cos k_x -\mu$,
$g(\vec{k})  = - 4 t^{\prime\prime} \sin k_x \sin k_y$, $\Delta_a(\vec{k}) =  \Delta \sin k_x \cos k_y$
and $\Delta_b(\vec{k}) = i \Delta \sin k_y \cos k_x$. The band parameters are chosen to be $(t,t^\perp,t^{\prime\prime},\mu)=(1,0.1,0.1,1.0)$~\cite{Raghu2010,Raghu2013}. $\Delta$ is the
overall superconducting gap magnitude, whose clean system value is left unspecified here since the normalized thermal conductivity $\kappa_0(\Gamma_N)/\kappa_0(\Gamma_c)$ to be calculated
does not depend on it. Because of the four-fold rotational crystal symmetry (preserved by our impurity scattering potential), there are
only four nonzero independent impurity self energy matrix elements: $\hat{\Sigma}_{11}=\hat{\Sigma}_{22}$, $\hat{\Sigma}_{12}=\hat{\Sigma}_{21}$, $\hat{\Sigma}_{33}=\hat{\Sigma}_{44}$,
and $\hat{\Sigma}_{34}=\hat{\Sigma}_{43}$. All other matrix elements are zero. Solving the coupled equations for these nonzero matrix elements as outlined previously and
calculating the thermal conductivity, we obtain the numerical result in Fig.~\ref{fig:kappaT}, the red $\diamondsuit$ points.
Comparing it to the single-band $d$-wave and isotropic chiral
$p$-wave pairing results we see that, interestingly, the results of $\kappa_0(\Gamma_N)/\kappa_0(\Gamma_c)$
for $\Gamma_N/\Gamma_c \gtrsim R \equiv |\Delta(\vec{k})|_{\min}/ |\Delta(\vec{k})|_{\max}\approx 7\%$ (see Appendix~\ref{app:gapprofile} Fig.~\ref{fig:Delta_P})
are almost identical to the $d$-wave pairing
result and can equally explain the experimental data points except the one at $\Gamma_N/\Gamma_c \approx 5\%$~\cite{Suzuki2002},
which is below the gap anisotropy ratio $R$.

If a smaller hybridization parameter $t^{\prime\prime}=0.05\, t$, defined in $g(\vec{k})$ of Eq.~\eqref{eq:H2band}, is taken as in Ref.~\onlinecite{Rozbicki2011},
then the gap minima in Appendix~\ref{app:gapprofile} Fig.~\ref{fig:Delta_P} become even deeper with a correspondingly smaller gap anisotropy ratio on the $\beta$ band,
$R\approx 3\%$. In this case all the experimental data points in Fig.~\ref{fig:kappaT} can be accounted for by the two-band model, as shown in the inset of Fig.~\ref{fig:kappaT}.

The fact that the agreement between the simple d-wave model used in Ref.~\onlinecite{Hassinger2017} and the 2-band chiral $p$-wave model for $\mathrm{Sr_2RuO_4}$
of Ref.~\onlinecite{Raghu2010} with $\Gamma_N\gtrsim \Delta_{\min}$ is surprisingly good with no adjustment of parameters merits some explanation.
It can be understood from the asymptotic expression of the residual thermal conductivity
\begin{gather} \label{eq:kapparatio-1}
\lim_{\Gamma_N\rightarrow 0} \frac{\kappa_0(\Gamma_N)}{\kappa_0(\Gamma_c)} =  \frac{2}{L_{\mathrm{FS}}} \sum_{i} \frac{\Gamma_c }{v_{\Delta}^i},
\end{gather}
where we have used $\lim_{\Gamma_N\rightarrow 0} \kappa_0(\Gamma_N)/(\pi^2 k_B^2 T/3)= (1/4\pi^2)\sum_iv_F/v_{\Delta}^i$
and $\kappa_0(\Gamma_c)/(\pi^2 k_B^2 T/3)=(L_{\mathrm{FS}} \, v_F /8\pi^2)1/\Gamma_c$ for a $2d$ single-band superconductor with a circular FS and
isotropic Fermi velocity $v_F$~\cite{Graf1996,Durst2000}.
Here $v_{\Delta}^i\equiv |\partial_{s} \Delta(s)|_{s=s_i}$ is the gap function slope along the FS contours, $s_i$ is the $i$-th node or near-node position on the FS,
and $L_{\mathrm{FS}}$ is the $2d$-FS contour length. For near-nodes, Eq.~\eqref{eq:kapparatio-1} is applicable only for $\Gamma_N\gtrsim \Delta_{\min}$.
Using $\Delta \cos(2\phi_{\vec{k}})$ on a circular FS and $\Gamma_c/\Delta=\sqrt{e}/4$ for
the single-band $d_{x^2-y^2}$ pairing~\cite{Sun1995} gives a value $=(2/\pi)\Gamma_c/\Delta\approx 0.26$ for the right hand side of
Eq.~\eqref{eq:kapparatio-1}~\cite{Sun1995}, in agreement with the blue line in Fig.~\ref{fig:kappaT}. Although Eq.~\eqref{eq:kapparatio-1}
is derived for a single-band superconductor, it can be applied to a $2$-band superconductor as well, provided that (1) the $v_F$ near the deep minima
is roughly the same as the averaged one over the entire FS and the averaged $v_F$ of different
bands is also roughly the same, which are the case for the $\alpha$ and $\beta$ bands of $\mathrm{Sr_2RuO_4}$~\cite{Mackenzie2003}, and (2) the total FS contour length
from all bands is used for $L_{\mathrm{FS}}$. In the $2$-band chiral $p$-wave model~\cite{Raghu2010}, only the $\beta$ band has very deep minima (see Appendix~\ref{app:gapprofile}
Fig.~\ref{fig:Delta_P}) and contributes to the sum of the right hand side of Eq.~\eqref{eq:kapparatio-1}; while both bands contribute to $L_{\mathrm{FS}}$, which
would reduce the ratio, $\lim_{\Gamma_N\rightarrow 0} \kappa_0(\Gamma_N)/\kappa_0(\Gamma_c)$, by about one half compared with that of the single-band $d_{x^2-y^2}$ pairing.
However, this reduction is compensated by the fact that the number of near-nodes on the $\beta$ band is $8$, double the number of nodes in the single-band $d$-wave case.
These cancelling factors of 2, and the fact that $\Gamma_c/v_{\Delta}^i$, after divided by $k_F$ to make it dimensionless, is comparable for the two models, account for
the agreement. A direct numerical evaluation of the right hand side of Eq.~\eqref{eq:kapparatio-1} for the $2$-band model gives a value $\approx 0.28$, consistent with
the red $\diamondsuit$ data in Fig.~\ref{fig:kappaT}.

We note that, in Ref.~\onlinecite{Hassinger2017}, the
experimental data was also compared to an SCTA result from Ref.~\onlinecite{Nomura2005}, which is obtained for a pairing model
with also extremely deep gap minima on the $\beta$ band. However, the residual thermal conductivity computed there is almost zero, in sharp contrast
to our results in Fig.~\ref{fig:kappaT} at $\Gamma_N/\Gamma_c\gtrsim \Delta_{\min}/\Delta_{\max}$.
The difference comes from the extremely small impurity concentration used in Ref.~\onlinecite{Nomura2005}, $n_i=10^{-6}$ per unit square of the lattice,
for which a rough estimate of $\Gamma_N$ for the $\gamma$ band, which has the largest gap in Ref.~\onlinecite{Nomura2005}, gives
$\Gamma_N=n_i/(\pi N_F) \approx 0.74 \, \mathrm{mK}$~\cite{Borkowski1994}. We have used the density of states of the
$\gamma$ band, $N_F\approx m_{\gamma}^* a^2/(2\pi \hbar^2)\approx 5 \; \mathrm{eV}^{-1}$ per unit lattice square, with $m^*_{\gamma}\approx 16 m_e$
and $a\approx 3.87 \AA$ the $\gamma$-band effective mass and the in-plane lattice constant of $\mathrm{Sr_2RuO_4}$, respectively~\cite{Mackenzie2003}.
This impurity scattering rate corresponds to $\Gamma_N/\Gamma_c\sim \Gamma_N/T_c\approx 5\times 10^{-4}$, if we use $T_c \approx 1.5 \mathrm{K}$.
This ratio is too small compared with the $\Gamma_N/\Gamma_c\approx 0.26$, estimated for the experimental sample in Ref.~\onlinecite{Hassinger2017}, which shows
that the calculation of Ref.~\onlinecite{Nomura2005} was performed in an extremely clean limit, and the
results obtained can not be directly compared to the experiment of Ref.~\onlinecite{Hassinger2017} at very low temperature (roughly speaking, not applicable when
$T/T_c\lesssim \Gamma_N/\Gamma_c$).

\subsection{$2d$ $3$-band model}
The second chiral $p$-wave pairing model we consider is the $2d$ $3$-band pairing model from Ref.~\onlinecite{Scaffidi2015} that was used to model the results
of a weak-coupling RG calculation~\cite{Scaffidi2014}. The gap structure on the $\alpha$ and $\beta$ bands is similar to the $2$-band model of Sec.~\ref{sec:SCTA_2band}, with
8 near-nodes on the $\beta$ band. This model fits critical specific heat jump data~\cite{Scaffidi2014} and has been used to explain the
absence of observable edge currents~\cite{Scaffidi2015}.

We choose the normal state part of $\hat{H}_{\mathrm{BdG}}$ to be identical to that from Refs.~\onlinecite{Scaffidi2014,Scaffidi2015}.
Following Ref.~\onlinecite{Scaffidi2015} we choose the superconducting order parameter matrix in the three orbital basis of $d_{xz},d_{yz}$ and $d_{xy}$ to be
$\hat{\Delta}_{\alpha,\beta}=\Delta_{\alpha}\delta_{\alpha,\beta}$, where $\Delta_{\alpha}$ for each orbital is a linear combination of different harmonics
consistent with chiral $p$-wave pairing on a square lattice:
\begin{align} \label{eq:3bandgap}
\begin{pmatrix}
\Delta_{xz} \\
\Delta_{yz} \\
\Delta_{xy}
\end{pmatrix}
& =
\begin{pmatrix}
a_1 & a_2  & a_3 \\
0 & 0 & 0 \\
b_1 & b_2 & b_3
\end{pmatrix}
\begin{pmatrix}
g_{1}^x \\
g_{2}^x \\
g_{3}^x
\end{pmatrix}
+ i
\begin{pmatrix}
0 & 0 & 0 \\
a_1 & a_2  & a_3\\
b_1 & b_2 & b_3
\end{pmatrix}
\begin{pmatrix}
g_{1}^y \\
g_{2}^y \\
g_{3}^y
\end{pmatrix}
\end{align}
where
\begin{subequations}\label{eq:ggg}
\begin{align}
(g_{1}^x,g_{2}^x,g_{3}^x) & =(\sin k_x, \sin k_x \cos k_y,\sin 3 k_x) \\
(g_{1}^y,g_{2}^y,g_{3}^y) & =(\sin k_y, \sin k_y \cos k_x,\sin 3 k_y).
\end{align}
\end{subequations}
Based on Ref.~\onlinecite{Scaffidi2015}, we choose the six coefficients $(a_1, a_2, a_3, b_1, b_2, b_3)=(0, 0.067, 0.33, 0.18, 0.15,-0.3)$
such that the gap functions on each band, obtained from $\Delta_{xz},\Delta_{yz}$ and $\Delta_{xy}$ by unitary transformation,
fit the weak coupling RG-calculation results well~\cite{Scaffidi2014}. This coefficient combination produces a ratio of the gap magnitude on
$\alpha,\beta$ bands to that on $\gamma$ band about $1/2$, which predicts a critical specific heat jump comparable to the experimental value~\cite{Scaffidi2014,Nishizaki2000}.
However, the pairing gap function obtained gives a result of $\kappa_0(\Gamma_N)/\kappa_0(\Gamma_c)$ quite different from that of the
$d$-wave case (see Appendix~\ref{app:gapprofile} Fig.~\ref{fig:kappaT2}) for two reasons:
\begin{enumerate}
\item The $\kappa_0(\Gamma_N)/T$ calculated, given in Fig.~\ref{fig:kappaT2} of Appendix~\ref{app:gapprofile}, has a multi-gap structure.
There is a peak at $\Gamma_N/\Gamma_c\approx 0.25$, which, however, is absent in the single band $d$-wave result and also not observed in $\mathrm{Sr_2 Ru O_4}$ measurements~\cite{Hassinger2017}.
That peak is a result of the $\alpha,\beta$ bands becoming normal, while the $\gamma$ band remains superconducting at $\Gamma_N/\Gamma_c\gtrsim 0.25$ (see
Appendix~\ref{app:gapprofile} for further discussions).\\

A more realistic model would include inter-band Cooper pair scattering that ensures a single $T_c$ for the three bands and, if sufficiently strong, may eliminate the peak at
$\Gamma_N/\Gamma_c\approx 0.25$. Here we avoid the multi-gap structure by simply adjusting the gap magnitude on each band so that the superconductivity on all three bands are destroyed at the same $\Gamma_N$. In other words, we impose a constraint on
the gap magnitude ratio among different bands. However, this constraint would likely be modified in the presence of interband interactions that are neglected
in our model.
\item Even if the gap magnitude ratio is adjusted such that the multi-gap structure disappears, the obtained
$\lim_{\Gamma_N\rightarrow 0} \kappa_0(\Gamma_N)/\kappa_0(\Gamma_c)$ value, when the deep minima are treated as accidental nodes, is still smaller than that of $d$-wave
(see Appendix~\ref{app:gapprofile} Fig.~\ref{fig:kappaT2}), which is about~\cite{Sun1995a} $0.26$.
This is not surprising since the $\alpha$ and $\beta$ bands alone would give approximately the $d$-wave value and the $\gamma$ band gap,
while anisotropic, does not have near-nodes.
From Eq.~\eqref{eq:kapparatio-1} we see that we need to either
decrease the gap function slope near deep minima or increase the number of deep minima. This is a serious constraint that the experimental data places on chiral $p$-wave
models with deep gap minima only on the $\alpha,\beta$ bands and not on the $\gamma$ band.
\end{enumerate}

In Appendix~\ref{app:gapprofile} we show that the $3$-band model with a reduced $v_{\Delta}$ at the $8$ near-nodes of the $\beta$ band agrees with the experimental
data. However, since the $v_{\Delta}$ is noticeably smaller than expected for $\mathrm{Sr_2RuO_4}$, it is useful to consider other ways one might reconcile
$3$-band chiral $p$-wave models with the residual thermal conductivity data. Models with horizontal nodes and with deeper minima on the $\gamma$ band are considered
below.

\subsection{$3d$ $3$-band model with horizontal line nodes} \label{sec:horizontal_node}
Horizontal line nodes in a $3d$ pairing model is one possibility for reconciling chiral $p$-wave with the residual thermal conductivity data.
Because of the highly quasi-$2d$ nature of $\mathrm{Sr_2RuO_4}$, which implies weak inter-layer coupling,
pairing with a strong $k_z$ dependence and, therefore, horizontal line nodes may seem unlikely, particularly on the $\gamma$ band which has the weakest interlayer coupling
~\cite{Bergemann2000,Mackenzie2003}.
However, chiral $p$-wave pairing models with either a $\cos k_z/2$ or $a+b \cos k_z$ dependence on $k_z$ (usually just on the
$\alpha$ and $\beta$ bands) have been proposed
to explain some experiments~\cite{Hasegawa2000,Zhitomirsky2001,Annett2002,Wysokinski2003,Litak2004,Koikegami2003,Kittaka2018}. Here $a$ and $b$ are two coefficients.
Note that higher order harmonics in $k_z$ in the gap function are more unlikely given the weak $k_z$ dependence of all three bands.

We can estimate the horizontal line node contribution to the residual thermal conductivity ratio using an analysis similar to that used to obtain
Eq.~\eqref{eq:kapparatio-1} together with values for the average Fermi velocities and lengths of Fermi surface in the $ab$-plane~\cite{Mackenzie2003}. Assuming horizontal line nodes on both
the $\alpha$ and $\beta$ bands at one or more values of $k_z$ we obtain
\begin{gather}
\lim_{\Gamma_N\rightarrow 0}\frac{\kappa_{0}(\Gamma_N)}{\kappa_{0}(\Gamma_c)}
\approx 0.23 \sum_{i} \frac{\Gamma_c \; c}{v_{\Delta,i}^{c}}. \label{eq:kapparatio}
\end{gather}
where $i$ is summed over the values of $k_z$ corresponding to horizontal nodes, $v_{\Delta,i}^{c}$
is the gap function velocity averaged over the horizontal line nodes of both the $\alpha$ and $\beta$ bands,
and we have restored the lattice spacing constant $c$ to make the expression explicitly dimensionless.
The superscript ``c" in $v^c_{\Delta,i}$ indicates that the gap velocity is along the $c$-axis direction. 
The case of accidental horizontal nodes (i. e., not protected by symmetry) can be modelled by a gap function $\Delta(k_z)\approx \Delta (a+b \cos k_z)/(|a|+|b|)$,
where $\Delta$ should be understood as the gap magnitude averaged over the in-plane FS contours of
both the $\alpha$ and $\beta$ bands and $|a|<|b|$.
In this case, from Eq.~\eqref{eq:kapparatio}
and using $\Gamma_c/\Delta=\sqrt{e}/4$ as a rough estimate~\cite{Sun1995}, one finds
\begin{gather}
\lim_{\Gamma_N\rightarrow 0}\frac{\kappa_{0}(\Gamma_N)}{\kappa_{0}(\Gamma_c)}\approx 0.19 \frac{|b|/|a|+1}{\sqrt{(b/a)^2-1}}. \label{eq:kapparatio-2}
\end{gather}
It follows that one could fit the experimental residual thermal conductivity with horizontal line nodes alone (without vertical nodes) if $b/a\sim 3.3$,
and we have confirmed this with a numerical calculation of
$\kappa_{0}(\Gamma_N)/\kappa_{0}(\Gamma_c)$.

From the above analysis, it is clear that models with both deep vertical minima and horizontal nodes on the $\alpha$ and $\beta$ bands~\cite{Annett2002,Wysokinski2003},
may be compatible with the experimental data depending on the details of these models. It follows from Eq.~\eqref{eq:kapparatio-2} and our previous numerical
results (Appendix~\ref{app:gapprofile} Fig.~\ref{fig:kappaT2}) that adding $a+b \cos k_z$
with $ b > a$ to the $\alpha$ and $\beta$ band pairing gap functions in Eq.~\eqref{eq:3bandgap}
and using the same coefficients in Eq.~\eqref{eq:3bandgap} as in Ref.~\onlinecite{Scaffidi2014,Scaffidi2015}
leads to a $\lim_{\Gamma_N\rightarrow 0}\kappa_{0}(\Gamma_N)/\kappa_{0}(\Gamma_c)$ value larger than that of the single-band $d$-wave.
The minimum of $\lim_{\Gamma_N\rightarrow 0}\kappa_{0}(\Gamma_N)/\kappa_{0}(\Gamma_c)$ is achieved when $a=0$,
which can be modelled by making the following replacement in Eq.~\eqref{eq:3bandgap}
\begin{equation} \label{eq:3dgap}
\{ g_1^x,g_2^x,g_1^y,g_2^y\} \rightarrow \{ g_1^x,g_2^x,g_1^y,g_2^y\} \times \cos k_z.
\end{equation}
However, even in this case, we will need to reduce $v_{\Delta}$ at the vertical deep minima slightly to fit the experimental data. Fig.~\ref{fig:kappaT} (dark-green $\times$) shows the residual
thermal conductivity for this model with both horizontal nodes and vertical near-nodes with parameters $(a_1,a_2,a_3,b_1,b_2,b_3)=(-0.1, 0.75, 2.0, 0.18, 0.15, -0.3)$.
Details of the gap function are given in Appendix~\ref{app:gapprofile} and, as before, the relative gap magnitudes of different bands have been tuned to vanish
at the same impurity concentration. Note, to explain the data with this model puts constraints on both the horizontal and vertical nodes or near-nodes.
Also, note that using $\cos k_z/2$ in Eq.~\eqref{eq:3dgap} is as good as $\cos k_z$, as can be seen from the previous estimates.

\subsection{$3$-band model with deep minima on the $\gamma$ band} \label{sec:SCTA_gamma}
While weak coupling RG calculations for $\mathrm{Sr_2RuO_4}$~\cite{Scaffidi2014} predict substantial anisotropy on the $\gamma$ band, the ratio of
minimum to maximum gap on the $\gamma$ band is predicted to be only $25\%$. However, functional RG (fRG) studies~\cite{Wang2013} found this ratio to be about $10\%$.
The deeper minima along the $k_x$ and $k_y$ axes may result from the fact that fRG  mixes in states away from the FS and closer to the Brillouin zone
boundary where the chiral $p$-wave gap must vanish by symmetry. However, this calculation also found much weaker superconductivity on the $\alpha/\beta$ bands, an effect
that may be modified if spin orbital coupling were to be included. In any case, the $3$-band functional RG results would give a poor fit to the experimental thermal
conductivity data because the superconductivity on the $\alpha/\beta$ bands is about an order of magnitude smaller than on the $\gamma$ band. Here, we combine the
fRG results for the $\gamma$ band with the simple model used in Sec.~\ref{sec:SCTA_2band} for the $\alpha/\beta$
bands.

The BdG Hamiltonian of the combined $3$-band model is
\begin{gather}\label{eq:H3band_gamma}
\hat{H}_{\mathrm{BdG}} =
 \begin{pmatrix}
  \epsilon_a        &         g         &          0   &        \Delta_a      & 0         & 0 \\
   g              &         \epsilon_b  &          0   &         0            &  \Delta_b  & 0 \\
   0                &          0          &    \epsilon_c &            0        & 0          & \Delta_c \\
  \Delta_a^*       & 0                   & 0              &  -\epsilon_a        & -g         & 0  \\
 0                 &   \Delta_b^*        & 0              &  -g                &  -\epsilon_b & 0 \\
 0                & 0                     & \Delta_c^*    & 0                   & 0           & -\epsilon_c
 \end{pmatrix},
\end{gather}
where for brevity we have suppressed the $\vec{k}$ dependence of all matrix elements. The definitions of $\epsilon_a(\vec{k}), \epsilon_b(\vec{k}), g(\vec{k}),
\Delta_a(\vec{k})$ and $\Delta_b(\vec{k})$ are identical to those given for Eq.~\eqref{eq:H2band}, except that the smaller orbital hybridization $t^{\prime\prime}=0.05\,t$
has been adopted here.
$\epsilon_c(\vec{k})=-2 t^\prime (\cos k_x +\cos k_y)
-4 t^{\prime\prime\prime} \cos k_x \cos k_y -\mu_c$, with $(t^\prime,t^{\prime\prime\prime},\mu_c)=(0.8, 0.35, 1.3)$, is the $\gamma$ band normal state energy dispersion,
taken from Ref.~\onlinecite{Wang2013}. The $\gamma$ band gap of Ref.~\onlinecite{Wang2013} can be approximated by
$\Delta_c(\vec{k})=b_1 (g_1^x + i g_1^y) + b_2 (g_2^x + i g_2^y) + b_3 (g_3^x+ i g_3^y)$, where $\{g_1^x,g_2^x,g_3^x,g_1^y,g_2^y,g_3^y\}$ are defined in Eq.~\eqref{eq:ggg}
and we choose the three coefficients to be $(b_1,b_2,b_3)=(-0.9,1,0.25)$. This functional form of $\Delta_c(\vec{k})$ gives an angular dependence of the $\gamma$
band gap function similar to the fRG results. In particular, the gap anisotropy ratio, $R=\Delta_{\min}/\Delta_{\max}\approx 10\%$, and the gap function slope near the deep minima on the $\gamma$ band FS, which are the two important things for the residual thermal conductivity
at small impurity scattering rate, are almost the same as in Ref.~\onlinecite{Wang2013}.
\footnote{In Ref.~\onlinecite{Wang2013}, the fRG gap has been approximated by the same $\Delta_c(\vec{k})$ but with the three coefficients $(b_1,b_2,b_3)=(-0.4375, 1, 0)$,
which, however, does not capture the smaller fRG gap function slope near the minima and, consequently, gives a residual thermal conductivity smaller
than the one calculated in Fig.~\ref{fig:kappaT}, black $\bigtriangleup$.}

The numerical results of $\kappa_0(\Gamma_N)/\kappa_0(\Gamma_c)$ are shown in Fig.~\ref{fig:kappaT} by the black $\bigtriangleup$.
Although the experimental data at $\Gamma_N/\Gamma_c \ge 15 \%$ can be accounted for by the combined $3$-band within experimental
error bars, the gap anisotropy ratio $R$ would need to be decreased such that  $R\lesssim 5\%$ to be consistent with the experimental data
point at $\Gamma_N/\Gamma_c\approx 5\%$ (assuming $v_{\Delta}$ and $v_F$ near the minima remain
the same). Therefore, the experimental data imposes quite severe constraints on the $3$-band model in the absence of horizontal nodes.

\section{Self-consistent BdG analysis} \label{sec:BdG}

\subsection{Model and parameters}
To study the nature of the low-energy states beyond the SCTA, including order parameter inhomogeneity, we self-consistently solve the real-space BdG equations in
the presence of dilute unitary impurities. We focus on the $2$-band chiral $p$-wave pairing model with deep minima from Ref.~\onlinecite{Raghu2010},
described in Sec.~\ref{sec:SCTA_results}. The BdG Hamiltonian for the case where $\vec{k}$ is a good quantum number is given by Eq.~\eqref{eq:H2band}. In this
section, we work in real space, where the BdG Hamiltonian on a square lattice is
\begin{align} \label{eq:H2bandReal}
H = &  -\sum_{i j,\alpha \beta,\sigma}  \Big[ t^{\alpha\beta}_{i, j} \; c^\dagger_{i \alpha \sigma} c_{j \beta \sigma} + \text{h.c.} \Big]
- \sum_{i, \alpha, \sigma} \mu_{i} \; c^\dagger_{i \alpha \sigma} c_{i \alpha \sigma} \nonumber  \\
& + \sum_{i j, \alpha, \sigma} \Big[ \Delta^{\alpha}_{i j,\sigma \overline{\sigma}} \; c^\dagger_{i \alpha \sigma} c^\dagger_{j \alpha \overline{\sigma}}
+ \text{h.c.} \Big]
\end{align}
where $c_{i \alpha \sigma}$ is the electron annihilation operator for site $i$.
As before, the orbital labels are $\alpha=a$ or $b$ (for $d_{xz}$ or $d_{yz}$ orbitals).
The model of Ref.~\onlinecite{Raghu2010} includes only nearest-neighbour and next-nearest-neighbour hopping and next-nearest neighbour pairing. The nonzero hopping matrix elements
are $t^{aa}_{i, i+\hat{x}}=t^{bb}_{i, i+\hat{y}} = t$,
$ t^{aa}_{i, i + \hat{y}} = t^{bb}_{i, i + \hat{x}} = t_{\perp}$ and
$t^{ab}_{i, i \pm \hat{x} \mp \hat{y}} = - t^{ab}_{i, i \pm \hat{x} \pm \hat{y}}=t''$,
where $(t,t_\perp,t^{\prime\prime})=(1.0,0.1,0.1)$. In the absence of disorder, the chemical potential $\mu_i=1$.
In this model, the chiral $p$-wave pairing order parameter is
$\Delta^{aa}_{i, i + \hat{x} \pm \hat{y};\sigma \overline{\sigma}} = - i \Delta_a$ (with $\Delta^{aa}_{i, i - \hat{x} \pm \hat{y}; \sigma \overline{\sigma}} = + i \Delta_a$)
and $\Delta^{bb}_{i, i \pm \hat{x} + \hat{y}; \sigma \overline{\sigma}} = + \Delta_b$ (with $\Delta^{bb}_{i, i \pm \hat{x} -\hat{y}; \sigma \overline{\sigma}} = - \Delta_b$)
with all other $\Delta_{ij,\sigma \overline{\sigma}}^\alpha=0$ and no inter-orbital pairing.
Here $\overline{\sigma} = - \sigma$ and we choose the spin quantization axis such that the spin part of the superconducting order parameter is in the triplet state
$\uparrow \downarrow +\downarrow \uparrow$.
$\Delta_a$ and $\Delta_b$ are functions of $i$ and $\boldsymbol{\delta}=\pm \widehat{x}\pm \widehat{y}$ in the presence of disorder.
In the uniform case this choice of pairing results in the chiral $p$-wave gap structure
\begin{equation}
\Delta(\textbf{k}) \approx 4 \big[  \Delta_a \; \text{sin}(k_x) \text{cos}(k_y) + i \Delta_b \; \text{sin}(k_y) \text{cos}(k_x) \big],
\end{equation}
whose real and imaginary parts correspond to $\Delta_a(\vec{k})$ and $\Delta_b(\vec{k})$ of Eq.~\eqref{eq:H2band}.

The BdG equations are solved together with the
self-consistent gap equation
\begin{equation}
\begin{split}
\Delta_{\alpha}(i,\boldsymbol{\delta})  = \frac{V}{2} \sum_{n = 1}^{ 2 N_x N_y }  \Big\{ u_{n \alpha }(i) v^*_{n \alpha}(i+\boldsymbol{\delta}) \\
- u_{n \alpha }(i+\boldsymbol{\delta}) v^*_{n \alpha}(i)  \Big\}  [ 1 - 2 f_n ]
\end{split}
\end{equation}
where $\alpha=\{a,b\}$, $f_n \equiv [ 1 + e^{E_n/T } ]^{-1}$.
$V$ is the attractive interaction strength and we use $V=2.5 t$. $N_x\times N_y$ is the square lattice size.
$(u_{n a}(i),u_{n b}(i); v_{n a}(i), v_{n b}(i))$ is the $n$-th Bogoliubov quasiparticle
wavefunction that diagonalizes the BdG Hamiltonian, with corresponding eigenvalue $E_n$, i.e.,
\begin{equation} \label{eq:Bogoliubov}
c_{i \alpha \sigma} = \sum_{n=1}^{2 N_x N_y} \Big[  u_{n \alpha}(i) \gamma_{n \sigma} +  v^*_{n \alpha}(i) \gamma^\dagger_{n \overline{\sigma}}  \Big],
\end{equation}
where $\gamma_{n\sigma}$ is the Bogoliubov quasiparticle annihilation operator.

We study the effect of dilute unitary scattering disorder through an on-site chemical potential term of strength $V^{\text{imp}}$ which is isotropic and diagonal
in the orbitals. $V^{\text{imp}}=100 t$ has been chosen for the unitary scattering limit. In the presence of disorder, the chemical potential is tuned such that
the electron density $\langle n \rangle \approx 2.6$ per site remains the same as that of the uniform and clean system with $\mu_i\equiv 1$.
For each disorder configuration we solve the BdG and gap equations self-consistently until the variational free energy, order parameter, and electron density are converged to within $\epsilon = 10^{-4}$.
At least ten disorder configurations are averaged over for the largest system sizes studied, $N_x = N_y = 100$, but the results do not qualitatively change for smaller systems, $N_x = N_y \leq 40$, with over 200 disorder configurations.

For clean system and with the above choice of parameters, the self-consistent gap in the band basis at $T = 0$ is
$\Delta_a=\Delta_b= 0.274$ with $\min\{\Delta_{\beta}(\vec{k})\} = 0.06$,
$\max\{\Delta_{\beta}(\vec{k})\} = 0.88$, $\min\{\Delta_{\alpha}(\vec{k})\} = 0.41$ and $\max\{\Delta_{\alpha}(\vec{k})\} = 1$ over the $\alpha$- and
$\beta$-sheets of Fermi surface (see Appendix~\ref{app:gapprofile} Fig.~\ref{fig:Delta_P}). Note, in order to reduce finite size effects in our numerical
BdG calculations, we have chosen a pairing interaction strength that corresponds to rather strong coupling, with a large gap.  Consequently, the density of impurities
at $\Gamma_c$ for our model is noticeably larger than that expected for $\mathrm{Sr_2RuO_4}$. However, other than changing the scale for disorder, this does not impact the low temperature
results that we show in this section.

To make a comparison with the experiment in Ref.~\onlinecite{Hassinger2017}, where the impurity scattering rate for the sample studied is estimated to be $\Gamma_N / \Gamma_c \approx 0.26$,
we need to estimate what impurity concentration, $n_{\text{exp}}$, that scattering rate corresponds to in our BdG calculation.
This can be done by using $n_{\text{exp}}/n_c = \Gamma_N / \Gamma_c$, where $n_c$ is the impurity density at which the disorder averaged order parameter vanishes.
For the parameters we have chosen, $n_c \approx 42\%$, which implies $n_{\text{exp}}\approx 11\%$.
Also, since our T-matrix calculation shows that the behavior of $\kappa_0(\Gamma_N)/\kappa_0(\Gamma_c)$ is quite different depending on whether $\Gamma_N/\Gamma_c\lesssim
R$ or $\Gamma_N/\Gamma_c\gtrsim R$, where $R=\Delta_{\text{min}}/\Delta_{\text{max}}\approx 7\%$ (see Appendix~\ref{app:gapprofile} Fig.~\ref{fig:Delta_P}), we focus on impurity concentrations around the value
$n_i^*\approx 3\%$ at which $n_i^*/n_c=\Gamma_N/\Gamma_c=R$.

\subsection{Inverse participation ratio}\label{sec:IPR}

\begin{figure}
 \centering
  \includegraphics[width=0.75\linewidth]{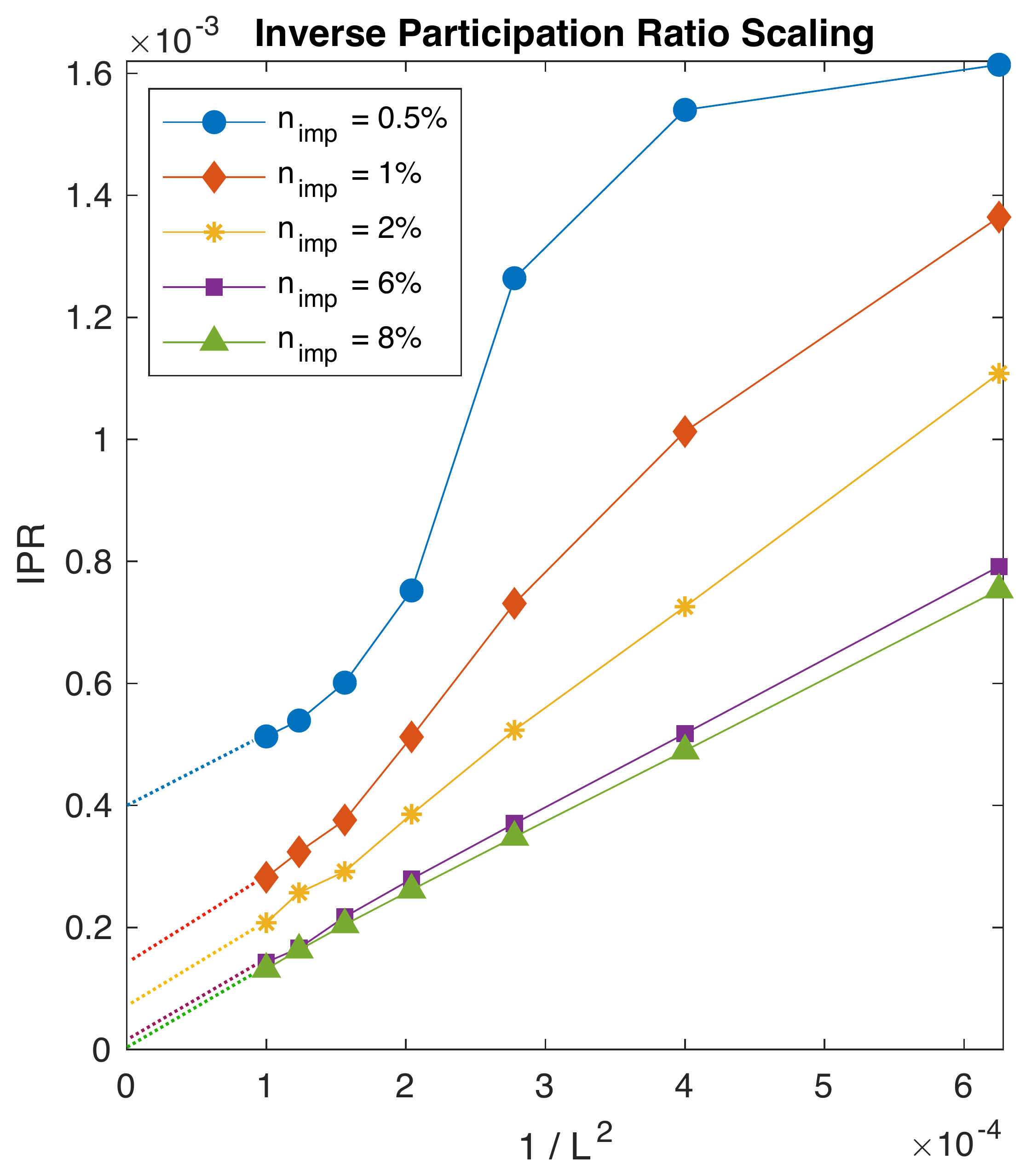}
  \caption{Scaling of the inverse participation ratio, averaged over states $n$ with energy $E_n\le \Delta_{\mathrm{min}}$, with $1/L^2$ for different
  impurity concentrations, where $L$ is the system size.
  For concentrations higher than $n_i^* \approx 3\%$, the IPR scales linearly and the extrapolated $\xi_L$ from the intercept is larger than the largest system size studied. The dashed lines
  are added to guide the eye. }
  \label{fig:IPR_P}
\end{figure}

The SCTA shows substantial residual thermal conductivity, implying that the low-energy states become delocalized, for scattering rates $\Gamma_N/\Gamma_c \gtrsim
\Delta_{\mathrm{min}}/\Delta_{\mathrm{max}}$ (for further discussions, see Appendix~\ref{app:localization_kappa}). Here we study the localized or non-localized nature of the low-lying states with varying disorder within self-consistent BdG,
by computing the inverse participation ration (IPR) $A_n$ given
by~\cite{Franz1996, Thouless1974}
\begin{equation}
A_n = \frac{ \langle | u_{n\alpha} |^4 \rangle + \langle | v_{n \alpha} |^4 \rangle }{ \Big[  \langle | u_{n \alpha} |^2 \rangle + \langle | v_{n \alpha} |^2 \rangle  \Big]^2 },
\end{equation}
where $\langle \cdots \rangle$ denotes sum over all sites and orbitals $\alpha=\{a,b\}$.  The IPR measures the reciprocal number of sites over which the quasiparticle wavefunction is delocalized,
and scales as $A_n \sim 1/L^2$ for extended states where $L$ is the system size.  For localized states, the IPR approaches $A_n \sim (a/\xi_L)^2$ as $L\rightarrow \infty$,
where $\xi_L$ is the characteristic localization length.\cite{Franz1996,Thouless1974}

In Fig. \ref{fig:IPR_P}, the IPR, $A_n$, averaged over states $n$ with
$E_n \leq \Delta_{\text{min}}$, is plotted versus $1/L^2$.  For the concentration $n_i=0.5\% < n_i^*\approx 3\%$,
the IPR extrapolates to a value corresponding to $\xi_L \leq 70$ lattice sites, whereas for concentrations $n_i > n_i^*$,
the IPR shows linear scaling with an extracted localization length greater than the largest system size studied. This shows that for $n_i\gtrsim n_i^*$,
or equivalently $\Gamma_N/\Gamma_c \gtrsim \Delta_{\text{min}} / \Delta_{\text{max}}$, the states near zero energy are delocalized and can make contributions
to the thermal transport, which supports our conclusion extracted from SCTA calculations on the residual thermal transport in Sec.~\ref{sec:SCTA}.
The existence of a threshold impurity concentration value, $n_i^*$, for sub-gap-minima states to be delocalized in the presence of deep gap minima
should be contrasted with the $d$-wave case. In that case the impurity-induced states mix with extended states and thereby
contribute to the thermal transport even with an infinitesimal amount of disorder, since the clean system has extended states all the way down to zero energy.

\subsection{Thermal Conductivity in BdG}\label{sec:BdGkappaT}

\begin{figure}
  \centering
  \includegraphics[width=0.75\linewidth]{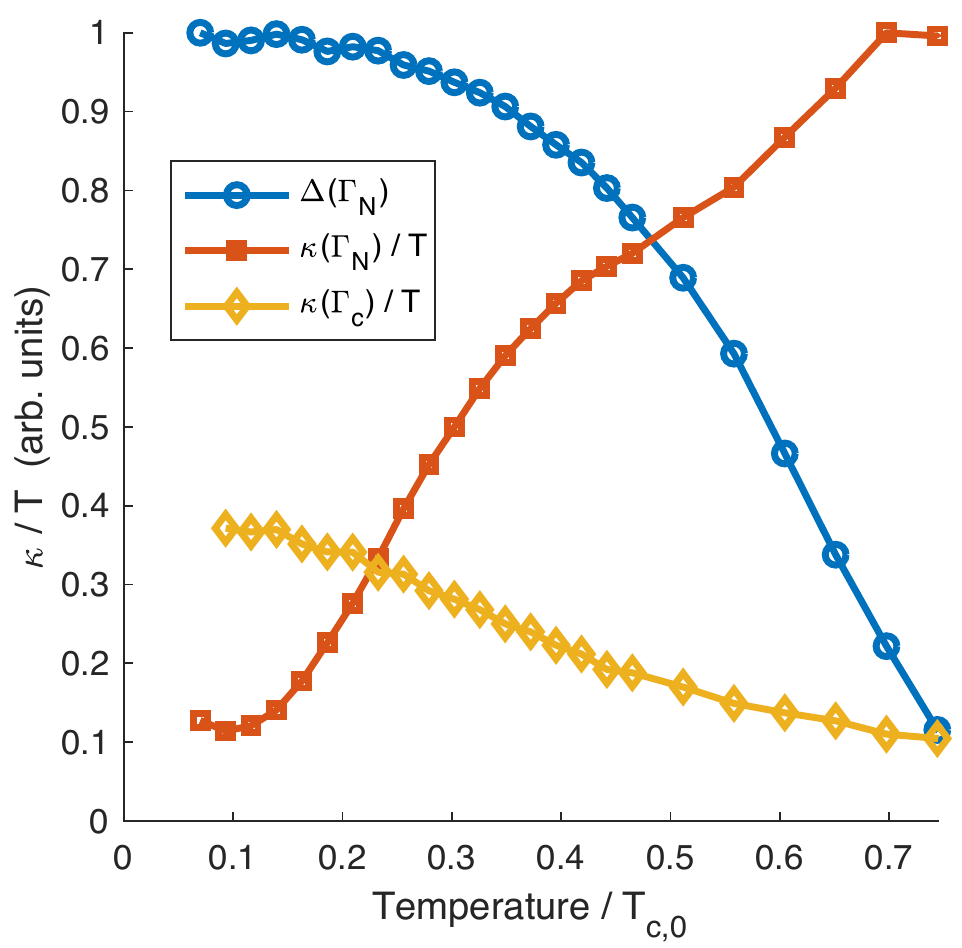}
  \caption{Thermal conductivity as a function of temperature computed from the self-consistent BdG.
  Blue circles correspond to the disorder-averaged gap $\langle \Delta \rangle$ scaled to its zero temperature value for impurity concentration
  $n_{\mathrm{exp}} = 11\%$.  Red squares and yellow diamonds are the $\kappa / T$ with $n_{\mathrm{exp}} = 11\%$ and $n_c = 42\%$, respectively.
  The two have the same units.
  All temperatures are scaled relative to the clean superconducting transition temperature $T_{c,0} \approx 0.43$.}
  \label{fig:Kappa_P}
\end{figure}

The longitudinal thermal conductivity can be computed from the Kubo formula
\begin{equation}
\kappa(T) = \frac{1}{T}  \lim_{\Omega \rightarrow 0} \frac{1}{\Omega}  \lim_{\vec{q} \rightarrow 0} \ \big\langle \text{Im} \ \Lambda_{x \, x}(\vec{q},
\Omega + i \delta) \big\rangle,
\end{equation}
where $\Lambda_{x \, x}$ is the $xx$-component of the thermal current-current correlation function tensor for a given disorder configuration and $\langle \cdots \rangle$ denotes
an average over different configurations.
Details of the calculation can be found in Appendix~\ref{app:kappa}. We show  the numerical results in Fig \ref{fig:Kappa_P} for two impurity
concentrations: $n_{\mathrm{exp}}\approx 11\%$, which corresponds to $\Gamma_N / \Gamma_c \approx 0.26$ from the experiment~\cite{Hassinger2017},
and the critical concentration $n_c\approx 42\%$. Quantities are plotted as a function of temperature relative to the clean transition temperature, $T_{c,0} \approx 0.43$.
The blue circles correspond to the disorder and spatially averaged gap $\langle \Delta \rangle$ scaled to its zero temperature value $\langle \Delta(T=0)\rangle$
for $n_{\mathrm{exp}} = 11\%$, which shows that the superconducting-to-normal transition occurs at $T_c/T_{c,0}\approx 0.8$. Note that the transition is significantly broadened by disorder.

The red squares in Fig.~\ref{fig:Kappa_P} are the $\kappa / T$ for $n_{\mathrm{exp}} = 11\%$, which shows a sizable residual $\kappa/ T$ at low
temperatures. This agrees with both our SCTA result and the experiment from Ref.~\onlinecite{Hassinger2017}.
We also calculate the $\kappa(T)/ T$ for the critical concentration $n_c \approx 42\%$, the yellow diamonds in Fig.~\ref{fig:Kappa_P}, from which the residual thermal
conductivity ratio can be extracted as roughly
$\kappa_0(\Gamma_N) / \kappa_0(\Gamma_c) \approx 0.32$. This ratio is also in good agreement with both the SCTA result in Fig.~\ref{fig:kappaT} and the
experiment~\cite{Hassinger2017}. However, note that this ratio is subject to numerical errors, because large fluctuations of the averaged gap magnitude in BdG
at large impurity concentrations makes an accurate determination of $n_c$ difficult.

While we do not perform a systematic study of the residual thermal conductivity dependence on scattering rate
within the BdG, due to the computational resources required for sufficient disorder averaging and larger system sizes, the rough estimate for this one
particular value of $\Gamma_N / \Gamma_c\approx 0.26$ suggests that the SCTA and BdG are in agreement and that our conclusions are valid beyond the approximations made
in treating the disorder scattering within the SCTA.

\section{Conclusions} \label{sec:conclusion}
Both the SCTA and self-consistent BdG calculations show that the residual thermal conductivity from deep gap minima or near-nodes in chiral $p$-wave
behaves similarly to that of $d$-wave provided $\Gamma_N/\Gamma_c\gtrsim \Delta_{\min}/\Delta_{\max}$ is satisfied. Although we have focused on chiral p-wave,
similar conclusions can be applied to other unconventional non-s-wave superconductors with deep minima. However,
our calculations illuminate the considerable constraints that the experimental thermal transport data~\cite{Suzuki2002,Hassinger2017} places on chiral $p$-wave pairing models
for $\mathrm{Sr_2RuO_4}$.
First, in order to account for all the experimental residual thermal conductivity data points in Fig.~\ref{fig:kappaT}, the gap minima
need to be sufficiently deep such that the gap anisotropy ratio satisfies $R\lesssim 5\%$~\cite{Suzuki2002}. However, this condition can have some caveats
since the experimental data points of Fig.~\ref{fig:kappaT} were obtained by assuming samples with $T_{c0}=1.5\mathrm{K}$ are in the true clean limit,
while a recent experiment~\cite{Kikugawa2016,Mackenzie2017} suggests that this might not be the case. If the $\Gamma_N/\Gamma_c$ for the cleanest sample in
Fig.~\ref{fig:kappaT} is higher than $5\%$, the condition $R\lesssim 5\%$ would be modified to a less-severe constraint.
Second, if there are no horizontal nodes, it is particularly difficult to reconcile chiral-$p$-wave order with residual thermal conductivity data.
While there are arguments for the possible existence of near-nodes on the $\alpha$ and $\beta$ bands, no similar arguments exist for the gap on the $\gamma$ band.
Weak coupling calculations that include spin orbital coupling do predict minima along the $(1,1)$ direction of the $ab$-plane for the gap on the $\gamma$ band, but these are not particularly
deep~\cite{Scaffidi2014}. On the other hand, minima along the $k_x$ or $k_y$ axes are expected by symmetry, but one would only expect these to be extremely deep if the FS was
extremely close to the zone boundary, in which case the contribution to the residual conductivity would become too small to explain the experimental data because of the
reduced $v_F$ near the zone boundary. We note that for vertical near-nodes along or near the $(1,1)$ direction, such as those on the $\alpha$ and $\beta$ bands, such
anisotropy of $v_F$ is not a concern because they are far away from the zone boundary.

The experimental data might be more easily accounted for by a $3d$ $3$-band chiral $p$-wave model with accidental horizontal line nodes or with a coexistence of vertical
near-nodes and horizontal line nodes. However, constraints exist even in such a model.
In the absence of horizontal nodes on the $\gamma$ band,
if two horizontal nodes exist on each of the $\alpha$ and $\beta$ bands alone without vertical near-nodes or other horizontal nodes, the gap velocity $v_{\Delta}$ at the horizontal nodes needs to be about $1/3$ of
that of simple $d$-wave to compensate for the absence of horizontal nodes on the $\gamma$ band; when accidental horizontal nodes on the $\alpha$ and $\beta$ bands coexist with vertical near-nodes,
the $v_{\Delta}$ at the horizontal nodes needs to be about $1/2$ of that of $d$-wave, depending on the $v_{\Delta}$ near
the vertical near-nodes.
We note that a recent $3d$ weak-coupling RG calculation~\cite{Roising2018} of the single-band repulsive Hubbard model found chiral $p$-wave order with horizontal line nodes
even when
the FS is a fairly weakly corrugated cylinder in the low electron density limit.
A similar $3d$ calculation for $\mathrm{Sr_2RuO_4}$, including all three bands and the $k_z$ dependence of the spin-orbital coupling~\cite{Haverkort2008}, would be very helpful to see
how favorable horizontal nodal gap structures are.

Lastly, we comment on some aspects of the thermal transport experiment~\cite{Hassinger2017} that we have left out in this study.
First, while we do not include magnetic fields in this paper, we expect that the residual thermal conductivity data at finite but small magnetic fields~
\cite{Hassinger2017} can be understood similarly as it only relies on quasiparticles excited near deep gap minima by the fields. Second,
the $c$-axis thermal transport also places considerable constraints on chiral $p$-wave models as the analysis in Ref.~\onlinecite{Hassinger2017} suggests that nodes or
near-nodes need to be present on all $3$ bands. This also emphasizes the importance of realistic microscopic $3d$ calculations for $\mathrm{Sr_2RuO_4}$.

\textit{Note added:} Recently, a $3$-band fRG calculation~\cite{Wang2018}, which takes into account the
spin-orbital coupling and finds extremely deep gap minima on the $\gamma$ band, has been reported by Wang \textit{et al}. They have calculated
the thermal conductivity at a finite temperature $T=T_{c}/30$ and compared the result to the experimental residual thermal conductivity data~\cite{Suzuki2002,Hassinger2017},
which is, however, obtained by extrapolating the finite $T$ data to $T=0$. Therefore, the quantity to be
compared with the experiments should be the one at $T=0$. Were the $T=0$ thermal conductivity used to compare
with the experiments in Ref.~\onlinecite{Wang2018}, the agreement would be poor at the smaller impurity scattering rates.
\section{Acknowledgments}
We would like to thank Andrew Millis, Louis Taillefer, Mark H. Fischer, and Steven A. Kivelson for discussions. This research is supported by the National Science and Engineering
Research Council of Canada (NSERC) (C. K. and Z. W.), the Canadian Institute for Advanced Research (CIFAR) (C. K. and Z. W.), and
the Department of Energy, Office of Basic Energy Sciences, under contract No. DE-AC02-76SF00515 (J. F. D.) at Stanford.
This work was made possible by the facilities of the Shared Hierarchical Academic Research Computing Network (SHARCNET:www.sharcnet.ca) and Compute/Calcul Canada.

\appendix \label{sec:appendix}
\begin{appendix}
\section{Gap function profiles and residual thermal conductivity} \label{app:gapprofile}
\begin{figure}[htp]
  \includegraphics[width=0.9\linewidth]{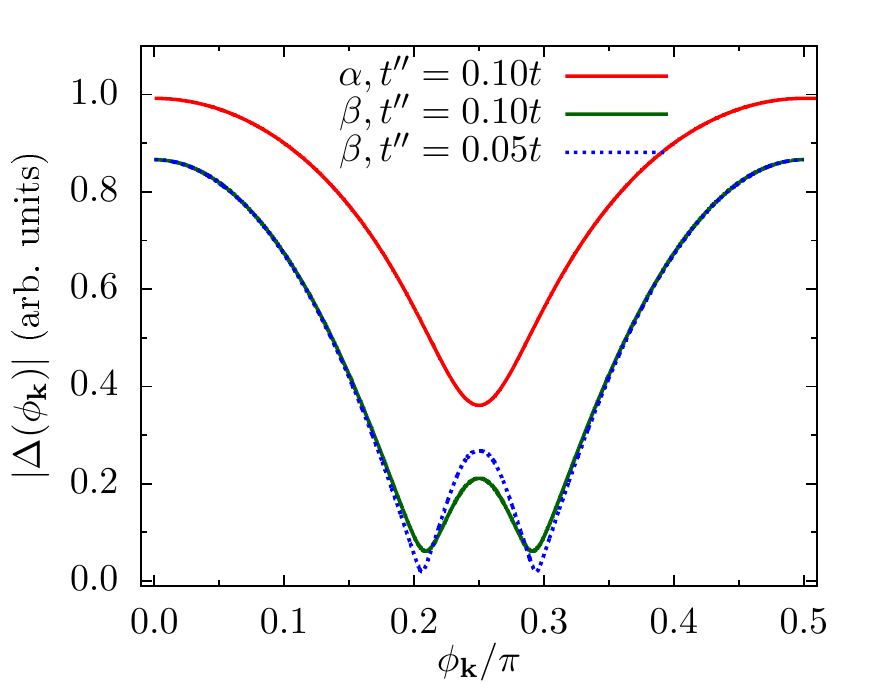}
  \caption{Gap magnitude, $|\Delta(\textbf{k})|$, of the $2$-band model on the $\alpha$ and $\beta$ FS sheets
  in the first quadrant of the $2d$ Brillouin zone. $\phi_{\vec{k}}$ is the azimuthal angle of Fermi wavevectors on each band and its definition
  can be found in the inset of Fig.~\ref{fig:3Bandv2Gap}. The top red and bottom dark-green solid lines are for the $\alpha$ and $\beta$ bands,
  respectively, obtained with the orbital hybridization parameter $t^{\prime\prime}=0.1\, t$ (see Eq.~\eqref{eq:H2band} for the definitions of $t$ and $t^{\prime\prime}$).
  For this parameter choice the gap anisotropy ratio on
  the $\beta$ band is $R=\Delta_{\min}/\Delta_{\max}\approx 7\%$. Also shown is the $\beta$ band gap function profile (bottom blue dotted line) with
  a smaller $t^{\prime\prime}=0.05 \, t$, for which $R\approx 3\%$.}
  \label{fig:Delta_P}
\end{figure}

\begin{figure}
\centering
\includegraphics[width=0.9\linewidth]{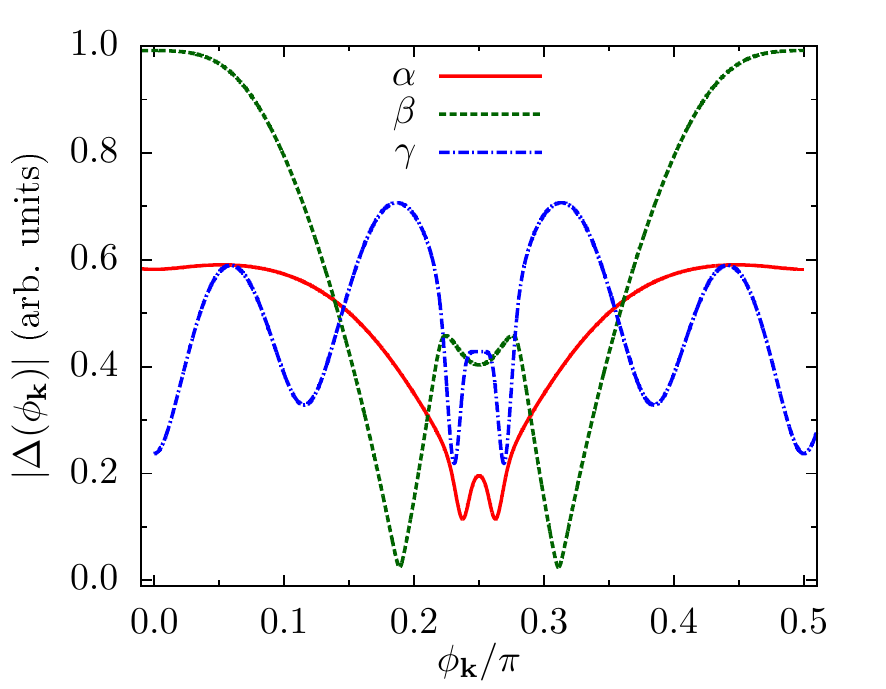}
\caption{Gap magnitude, $|\Delta(\phi_\vec{k})|$, of the $3d$ $3$-band pairing model, defined in Eq.~\eqref{eq:3dgap}, along the three FS sheets at $k_z=0$.
The FS contours at $k_z=0$ and the definition of the azimuthal angle, $\phi_{\vec{k}}$, are the same as in the inset of Fig.~\ref{fig:3Bandv2Gap}.}
\label{fig:3dgapkz0}
\end{figure}

\begin{figure}
\centering
\includegraphics[width=0.9\linewidth]{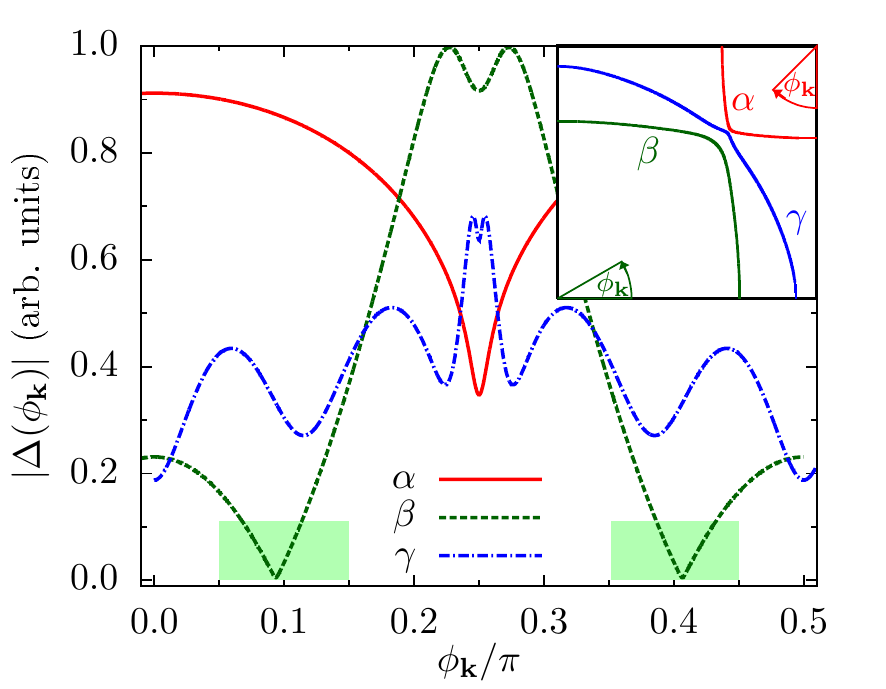}
\caption{Gap magnitude, $|\Delta(\phi_\vec{k})|$, of the $2d$ $3$-band model, used in
Fig.~\ref{fig:kappaT2}, along the three FS contours in the first quadrant of the $2d$ Brillouin zone. $|\Delta(\phi_{\vec{k}})|$ is symmetric with respect to
$\phi_{\vec{k}}/\pi=1/4$ on each band. The inset shows the corresponding FS
contours. $\phi_{\vec{k}}$ is the azimuthal
angle of Fermi wavevectors on each band, defined with respect to the center of each Fermi surface sheet. The key feature for the residual thermal conductivity is the
reduced gap slopes at the accidental nodes of the $\beta$ band, which is highlighted by the shaded areas in dark-green. }
\label{fig:3Bandv2Gap}
\end{figure}

In this Appendix we show the gap function profiles of the chiral $p$-wave models used in Fig.~\ref{fig:kappaT}, along the Fermi surface contours in the $k_x-k_y$ plane.
Fig.~\ref{fig:Delta_P} is for the $2$-band model, defined in Eq.~\eqref{eq:H2band},
and Fig.~\ref{fig:3dgapkz0} is for the $3d$ $3$-band model (defined in Eq.~\eqref{eq:3dgap}) in the $k_z=0$ plane. Fig.~\ref{fig:3Bandv2Gap} shows a modified $2d$ $3$-band gap function
profile that is discussed below.

From Fig.~\ref{fig:3dgapkz0} we see that the overall gap function profiles of our $3d$ $3$-band pairing model at $k_z=0$ are similar to those from
Refs.~\onlinecite{Scaffidi2014,Scaffidi2015}. This model supports horizontal line nodes at $k_z=\pm \pi/2$, which add a considerable contribution to the residual thermal conductivity and compensate for the fact that there are no deep vertical minima on the $\gamma$
band.

\begin{figure}
\centering
\includegraphics[width=0.9\linewidth]{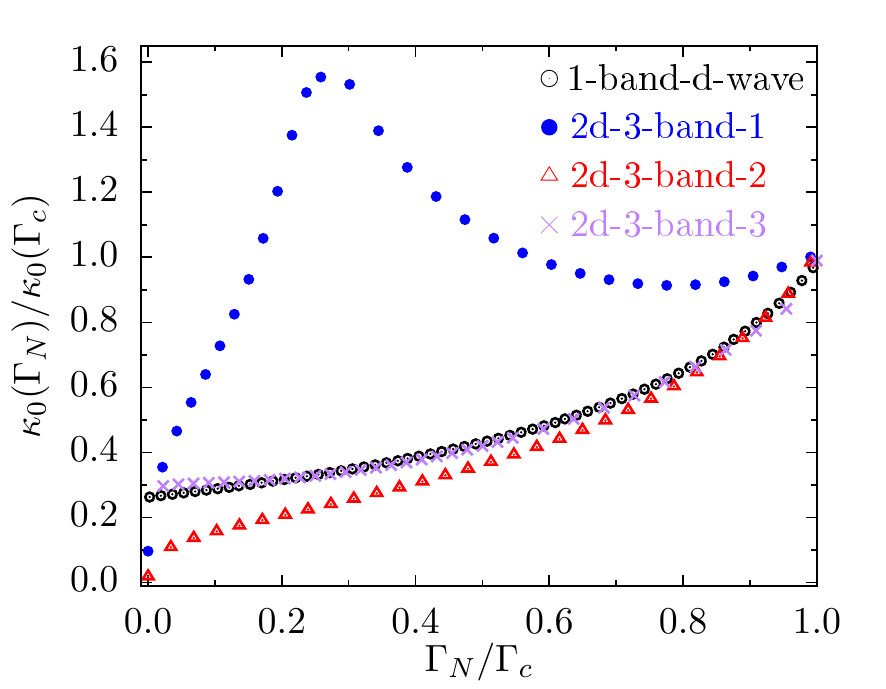}
\caption{$\kappa_0(\Gamma_N)/\kappa_0(\Gamma_c)$ calculated for the 2d $3$-band chiral $p$-wave pairing models (blue filled circles, red triangles, and purple crosses) defined in Eq.~\eqref{eq:3bandgap}.
The data labelled by ``2d-3-band-1" is obtained using the fitting gap parameters from Refs.~\onlinecite{Scaffidi2014,Scaffidi2015}\ ; while the ``2d-3-band-2"
is obtained by using the same angular dependence of the gap functions as in Refs.~\onlinecite{Scaffidi2014,Scaffidi2015} but adjusting the relative
gap magnitudes on different bands to eliminate the multi-gap structure. Both are different from that of the single band $d_{x^2-y^2}$-wave result (black open circles).
The `2d-3-band-3" is obtained by using a model with $v_{\Delta}$
reduced by about $1/2$ (see text) compared with that in Refs.~\onlinecite{Scaffidi2014,Scaffidi2015} and is similar to the single-band $d$-wave result. }
\label{fig:kappaT2}
\end{figure}

Fig.~\ref{fig:kappaT2} shows the $\kappa_0(\Gamma_N)/T$ calculated within the SCTA for the $2d$ $3$-band pairing model defined in Eq.~\eqref{eq:3bandgap}.
The blue filled circle is obtained by using the gap parameters from Refs.~\onlinecite{Scaffidi2014,Scaffidi2015}.
The $\Gamma_N$ dependence of $\kappa_0(\Gamma_N)/\kappa_0(\Gamma_c)$ in Fig.~\ref{fig:kappaT2} is quite different from that of the single-band $d$-wave because of the peak
near $\Gamma_N/\Gamma_c\approx 0.25$, which comes from the $\alpha$ and $\beta$ bands becoming normal while the $\gamma$ band remains superconducting at
$\Gamma_N/\Gamma_c\gtrsim 0.25$.

In our calculation, the $\alpha$ and $\beta$ bands are coupled and undergo the superconducting-to-normal
transition at the same critical impurity concentration, $n_c^{\alpha,\beta}$; on the other hand, the $\gamma$ band is almost uncoupled to the $\alpha$
and $\beta$ bands, and therefore has a different critical impurity concentration, $n_{c}^{\gamma}$. The values of $n_c^{\alpha,\beta}$ and $n_{c}^{\gamma}$ are determined
by $\Gamma_N=n_i/(\pi N_F) \sim T_c \propto \Delta$ for each band, where $N_F$ is the corresponding normal state density of states at Fermi energy.
This leads to $n_c^{\alpha,\beta}/n_{c}^{\gamma}\approx (N_F^{\beta}/N_F^{\gamma}) (\Delta^{\beta}/\Delta^{\gamma})\approx 0.25$, where we have used,
in determining $n_{c}^{\alpha,\beta}$, that the $\beta$ band dominates because it has a larger density of states~\cite{Mackenzie2003},
$N_F^{\beta}/N_F^{\gamma}\approx 0.63$~\cite{Mackenzie2003} and $\Delta^{\beta}/\Delta^{\gamma}\approx 0.4$
in the chiral $p$-wave model of Ref.~\onlinecite{Scaffidi2014} (see Fig. 4(a) there).
This explains why a peak occurs at $\Gamma_N/\Gamma_c\approx 0.25$ in Fig.~\ref{fig:kappaT2}.

Similar multi-gap structures in specific heat, London penetration depth, and thermal conductivity as a function of temperature have been predicted before, such as in
Ref.~\onlinecite{Agterberg1997}. However, neither those predictions nor the peak at $\Gamma_N/\Gamma_c\approx 0.25$ in Fig.~\ref{fig:kappaT2}
have been observed in experiments~\cite{Nishizaki2000,Bonalde2000,Suzuki2002,Hassinger2017}
\footnote{In Ref.~\onlinecite{Nishizaki2000}, a shoulder anomaly was observed in the low-temperature and in-plane magnetic field dependent specific heat, $C_{e}(H)$,
at $H\approx 0.1 H_{c2}^{ab}$, where $H_{c2}^{ab}$ is the in-plane upper critical field. This was suggested to support a two-gap scenario with the anomaly
attributed to the suppression of the minor gap. However, the more recent specific heat measurement from Ref.~\onlinecite{Kittaka2018} suggests that the shoulder
anomaly in Ref.~\onlinecite{Nishizaki2000} is due to an insufficient subtraction of the non-electronic contribution to the total specific heat, and the two-gap structure
disappears after a more complete subtraction.}, which, in our model, constrains the ratio of the superconducting gap magnitude on
the $\alpha$ and $\beta$ bands to that on the $\gamma$ band to be larger than $\Delta^{\beta}/\Delta^{\gamma}\approx 0.4$.
As emphasized in the main text, we neglect possible inter-band Cooper pair scattering
in our simplified model, and, therefore, this is not necessarily an actual constraint on the gap ratios in $\mathrm{Sr_2RuO_4}$.

To eliminate the multi-gap structure we simply adjust the relative gap magnitudes among different bands such that they vanish at the same $\Gamma_N$ but keep the angular dependence of
the gap functions on each band the same as in Refs.~\onlinecite{Scaffidi2014,Scaffidi2015}. The calculated  $\kappa_0(\Gamma_N)/T$ is shown in Fig.~\ref{fig:kappaT2}
by the red $\bigtriangleup$. We see that, similar to the results presented in Fig.~\ref{fig:kappaT}, the $\kappa_0(\Gamma_N)/\kappa_0(\Gamma_c)$ appears to saturate to a nonzero constant
as $\Gamma_N/\Gamma_c\rightarrow 0$, provided that $\Gamma_N/\Gamma_c\gtrsim \Delta_{\min}/\Delta_{\max}$. However, that constant is only about $1/2$ of that of
$d$-wave. From Eq.~\eqref{eq:kapparatio-1} we see that,
in order to increase the value of $\lim_{\Gamma_N\rightarrow0}\kappa_0(\Gamma_N)/\kappa_0(\Gamma_c)$ (treating the near-nodes as accidental nodes),
we need to either increase the number of near-nodes or
decrease the gap slope $v_{\Delta}$ at each near-node by about $1/2$. This puts a strong constraint on the possible gap function profiles in pairing models
with the near-nodes existing only on the $\alpha$ and $\beta$ bands and not on the $\gamma$ band.

We choose to reduce $v_{\Delta}$ by about $1/2$ by choosing the six coefficients in Eq.~\eqref{eq:3bandgap} to be
$(a_1,a_2,a_3,b_1,b_2,b_3)\propto (-0.8,0.4,1.6,0.18,0.15,-0.3)$.
As a consequence of the simple parameterization we are using, the near-nodes on the $\beta$ band (which are actually accidental gap nodes for the chosen parameter)
are inevitably shifted away from the zone diagonal (see Fig.~\ref{fig:3Bandv2Gap}).
We adjust the magnitudes of the gap functions in the band basis such that the superconductivity on all three bands vanishes at the same $\Gamma_N$.
Using the gap functions obtained in this way, we calculate the impurity self-energy matrix $\hat{\Sigma}$ and residual thermal conductivity,
following the procedure that we outlined in Sec.~\ref{sec:SCTA} of the main text.
Due to the four-fold rotational symmetry between $d_{xz}$ and $d_{yz}$ orbitals, the impurity self energy matrix elements satisfy:
$\hat{\Sigma}_{11}=\hat{\Sigma}_{22}$, $\hat{\Sigma}_{12}=\hat{\Sigma}_{21}$, $\hat{\Sigma}_{44}=\hat{\Sigma}_{55}$, $\hat{\Sigma}_{45}=\hat{\Sigma}_{54}$.
All other matrix elements are zero except $\hat{\Sigma}_{33}$ and $\hat{\Sigma}_{66}$ because of the orthogonality between $\big\{ d_{xz},d_{yz} \big\}$ and $d_{xy}$
orbitals. The numerical result of $\kappa_0/T$ is given in Fig.~\ref{fig:kappaT2} by the purple $\times$ points.
We see that they are similar to the $d$-wave results and can account for the experimental data.
However, this fit did require a $v_{\Delta}$ that is significantly smaller than the weak coupling RG results predict and is unlikely to
be realized in $\mathrm{Sr_2RuO_4}$.

\section{Localization effects on the residual thermal conductivity within SCTA} \label{app:localization_kappa}
Although, in the SCTA method, the translational invariance is restored in real-space after impurity-averaging, which seems to imply the underlying
states being extended, the signature of localization on transport quantities, such as the residual thermal conductivity, can still appear~\cite{Joynt1997}.

For illustration, we consider a quasi-$2d$ single band superconductor with no $k_z$ dependence, and assume an isotropic Fermi surface as well as a $|\vec{k}|$ independent
non-$s$ wave order parameter, $\Delta(\phi_{\vec{k}})$, where $\phi_{\vec{k}}$ is the azimuthal angle of $\vec{k}$. The residual thermal conductivity can be calculated from Eq.~\eqref{eq:kappa0}.
After an integration along the direction perpendicular to the circular Fermi surface we get~\cite{Ambegaokar1964,Ambegaokar1965,Maki1999,Nomura2005}
\begin{gather}
\frac{\kappa_0(\Gamma_N)}{T} \propto \bigg\langle \frac{1}{\Gamma_s(\omega=0,\phi_{\vec{k}})}
\bigg\{ 1+C(\omega=0,\phi_{\vec{k}}) \bigg\} \bigg\rangle_{\phi_{\vec{k}}}. \label{eq:kappa3}
\end{gather}
Here $\Gamma_s(\omega,\phi_{\vec{k}})\equiv \sqrt{(\mathrm{Im}\Sigma_0)^2+|\Delta(\phi_{\vec{k}})|^2}$ is the effective (frequency-dependent) impurity scattering rate of the superconducting state~\cite{Kadanoff1964}. If $\Delta(\phi_{\vec{k}})\equiv 0$
it reduces to the normal state impurity scattering rate, $\Gamma_N$. $\mathrm{Im} \Sigma_0$ is the imaginary part of
the diagonal impurity self-energy in the Nambu particle-hole space at $\omega=0$, and it depends on $\Gamma_N$. In Eq.~\eqref{eq:kappa3},
\begin{gather}\label{eq:oneplusC}
1+C(\omega=0,\phi_{\vec{k}})= \frac{2 \, (\mathrm{Im}\Sigma_0)^2}{(\mathrm{Im}\Sigma_0)^2+|\Delta(\phi_{\vec{k}})|^2}.
\end{gather}
$C(\omega,\phi_{\vec{k}})$ has been called a coherence factor in the literature, such as in Refs.~\onlinecite{Kadanoff1964,Nam1967}, which, however,
should not be confused with the usual coherence factors constructed from eigenfunctions of a BdG Hamiltonian~\cite{Schrieffer1999}.

In Eq.~\eqref{eq:kappa3}, it is precisely the $1+C$ factor that
gives the large difference between the $d$-wave and isotropic chiral $p$-wave results
in Fig.~\ref{fig:kappaT} at small $\Gamma_N$. In the former case, $\kappa_0(\Gamma_N)/\kappa_0(\Gamma_c)$ saturates to a nonzero constant as $\Gamma_N\rightarrow 0$,
while in the latter, it becomes vanishingly small.
In the $d$-wave case, the $\phi_{\vec{k}}$ average in Eq.~\eqref{eq:kappa3} mainly comes from the $\phi_{\vec{k}}$ regime near the nodes where
$|\Delta(\phi_{\vec{k}})|\lesssim \mathrm{Im}\Sigma_0$. As a consequence, $1\lesssim 1+C\le 2$ and it does not play a significant role. On the other hand, for
the isotropic chiral $p$-wave, $|\Delta(\phi_{\vec{k}})|\equiv \Delta=const.$ and $1+C \approx 2  (\mathrm{Im}\Sigma_0)^2/|\Delta|^2 \rightarrow 0$ as
$\Gamma_N/\Gamma_c \rightarrow 0$. This additional dependence on $\mathrm{Im} \Sigma_0$ makes the $\kappa_0/T$ vanish as
$\Gamma_N\rightarrow 0$ for the isotropic chiral $p$-wave, unlike for the $d$-wave, even though the impurity induced density of states, $N(\omega=0)$, rises rapidly with $\Gamma_N$
in both cases~\cite{Durst2000,Sun1995,Maki1999}, i.e., $N(0)\propto \mathrm{Im}\Sigma_0 \propto \sqrt{\Gamma_N \Delta}$ at small $\Gamma_N$,
where $\Delta$ is the clean system gap magnitude (we have ignored a logarithmic correction to $N(0)$ for the $d$-wave).

If the pairing is an anisotropic chiral $p$-wave with deep minima, $\Delta_{\min}$, then the behavior of $\kappa_0/T$ can be similar to either the $d$-wave
or the isotropic chiral $p$-wave case, depending on whether $\mathrm{Im}\Sigma_0 \gtrsim \Delta_{\min}$ or not in Eq.~\eqref{eq:oneplusC}. If we use
$\mathrm{Im}\Sigma_0 \propto \sqrt{\Gamma_N \Delta}$, then the $d$-wave and isotropic chiral $p$-wave like regimes
are delineated by $\Gamma_N\sim \Delta^2_{\min}/\Delta$. In other words, when $\Gamma_N \gtrsim \Delta_{\min}> \Delta^2_{\min}/\Delta$ (as a conservative condition),
we expect a $\kappa_0/T$ behavior similar to that of the $d$-wave. Although the results here are obtained for a single band superconductor, similar conclusions
hold for the multi-band pairing models that we considered in Fig.~\ref{fig:kappaT}.

Since the residual thermal conductivity $\kappa_0/T$ comes from the non-interacting Bogoliubov quasiparticle states at zero energy,
we can write $\kappa_0/T$ as~\cite{AshcroftMermin}
\begin{gather}\label{eq:kappaT5}
\kappa_0/T \propto (C_v/T) \, v^2 \, \tau,
\end{gather}
where $C_v/T \propto N(0)$ is the specific heat coefficient, $v^2$ is the mean square velocity of the Bogoliubov quasiparticles,
and $\tau$ is their effective mean free time. In the isotropic chiral $p$-wave case, from  Eq.~\eqref{eq:kappa3}, $\kappa_0/T \propto (\mathrm{Im}\Sigma_0)^2
\propto \Gamma_N$ at small $\Gamma_N$. Taking $v=v_F$~\footnote{ More precisely, the velocity here should be the group velocity of
Bogoliubov quasiparticles~\cite{Durst2000}. However, this does not affect our qualitative discussions.} and using $N(0)\propto \sqrt{\Gamma_N \Delta}$ we reach the conclusion that
$\tau \propto \sqrt{\Gamma_N \Delta} \rightarrow 0$ as $\Gamma_N\rightarrow 0$, which implies localized zero energy Bogoliubov states induced by disorder.
This is consistent with the single impurity result that a potential scatterer can induce Andreev bound states well below the isotropic
chiral $p$-wave gap (similar to the conclusion reached for an $s$-wave superconductor with a paramagnetic impurity~\cite{Balatsky2006}).
In Eq.~\eqref{eq:kappa3}, the Andreev bound state nature of the zero energy states
is reflected in the $1+C$ factor, which is why it makes a big difference between the isotropic chiral $p$-wave and $d$-wave.

A similar interpretation can be made for the $d$-wave case and leads to $\tau \propto 1/\sqrt{\Gamma_N\Delta} \rightarrow \infty$
as $\Gamma_N\rightarrow 0$, indicating delocalized zero energy states, which, again, agrees with the single impurity result~\cite{Balatsky2006},
although the localization issue of this case is quite subtle~\cite{Balatsky2006}.

\section{Disorder effects on the low energy density of states of $s$- and non-$s$-wave superconductors with accidental nodes or near-nodes} \label{app:DOS}
The different effects of disorder on accidental nodes or near-nodes in non-$s$-wave and $s$-wave superconductors are easily seen in the SCTA.
Here, we show this difference in self-consistent BdG calculation of the disorder averaged density of states (DOS) at low energy.

\begin{figure}
\centering
  \includegraphics[width=0.65\linewidth]{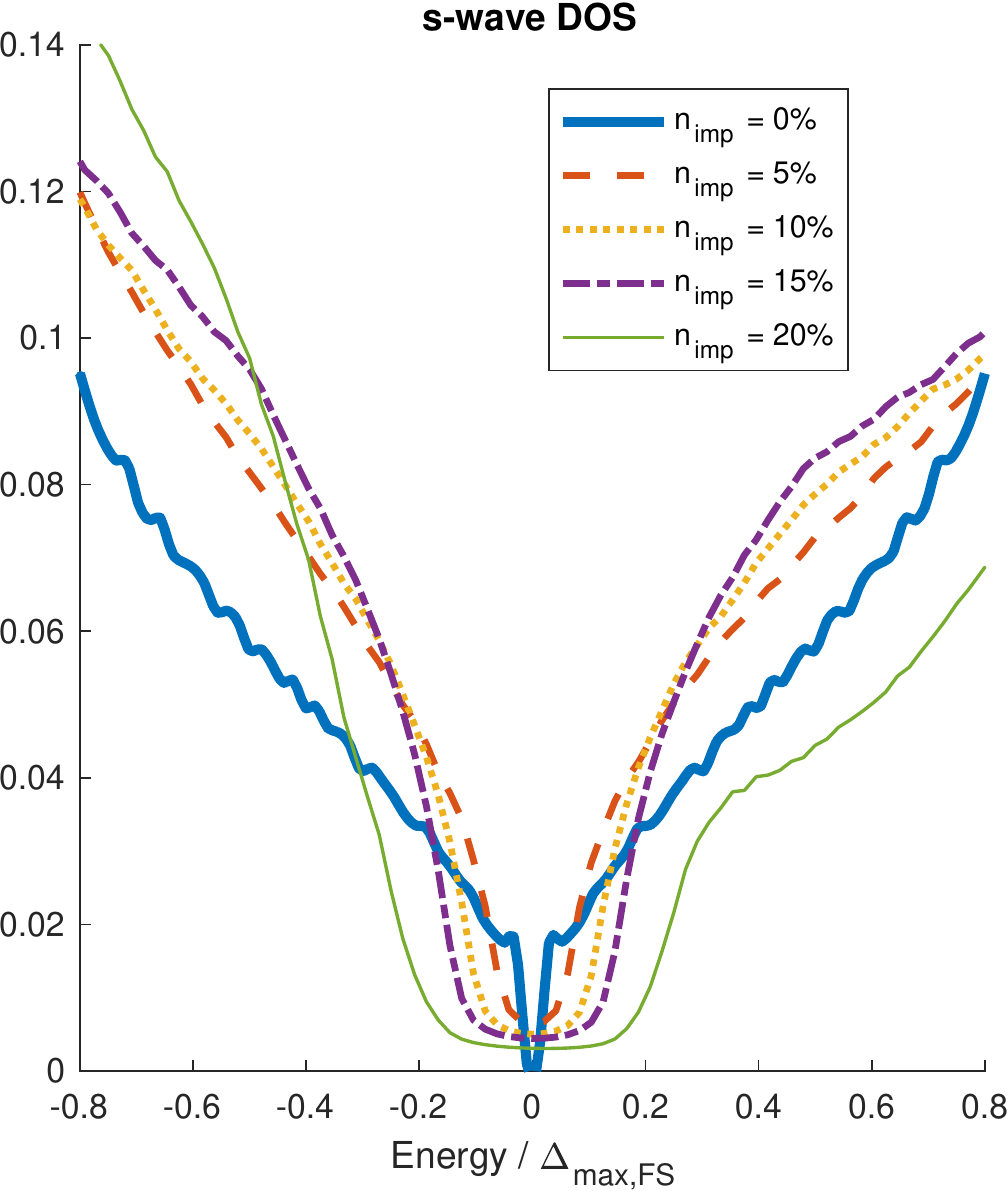}
  \caption{Disorder-averaged DOS at low energy for the anisotropic $s$-wave model of
  Eq.~\eqref{eq:swave} with impurity concentrations $n_{\text{imp}} = 0, 5, 10, 15, 20\%$ from self-consistent BdG calculations. The energy is normalized to the clean
  system maximum gap on the FS,
  while the DOS here and the one in Fig.~\ref{fig:DOS_P} are normalized to unity over the normal state bandwidth.}
  \label{fig:DOS_S}
\end{figure}

In an $s$-wave superconductor, non-magnetic impurity scattering neither induces states well below the minimum gap
nor changes the $\vec{k}$-space averaged gap magnitude, $\overline{\Delta(\vec{k})}$, within the SCTA.~\cite{Balatsky2006} 
However, the difference of the gap magnitude from $\overline{\Delta(\vec{k})}$ at each $\vec{k}$ is renormalized by the disorder~\cite{Balatsky2006} and
it decreases as the disorder increases, implying that the gap minima increase with disorder. This is indeed seen in Fig.~\ref{fig:DOS_S}, where we plot the
disorder averaged DOS at different impurity concentrations, computed from self-consistent BdG for a single band $s$-wave superconductor with the
following pairing gap function~\cite{Borkowski1994}
\begin{equation}\label{eq:swave}
\Delta(\vec{k}) =  \Delta_0 \big[ 1 - \text{cos}(k_x+k_y)  \big] \big[ 1 - \text{cos}(k_x - k_y) \big].
\end{equation}
$|\Delta(\vec{k})|$ has the same nodal structure as $d$-wave along the diagonal, but $\Delta(\vec{k})$ does not change sign
near the nodes. The nodes are accidental rather than symmetry enforced. From Fig.~\ref{fig:DOS_S} we see that the roughly linear DOS at low energy in the clean system gives way
to a gap which grows with increasing disorder. Similar results have been observed in the SCTA~\cite{Borkowski1994}.
However, our self-consistent BdG result in Fig.~\ref{fig:DOS_S} shows that the above conclusion about gap anisotropy holds even beyond the SCTA at very high impurity concentrations, where the spatial
variations of the local order parameter become important~\cite{Ghosal1998,Trivedi2001}
\footnote{From the inverse participation ratio we find that the low energy states at the band bottom of Fig.~\ref{fig:DOS_S} are localized when the impurity density
is sufficiently large, suggesting a transition from the superconducting to a gapped insulating phase at large disorder, which is related to the breakup of the system
into superconducting islands separated by non-superconducting sea. The results are qualitatively similar to those obtained in previous studies of isotropic $s$-wave superconductors with random chemical potential disorder\cite{Ghosal1998,Trivedi2001}.}
and the SCTA is not applicable.

\begin{figure}
\centering
  \includegraphics[width=0.65\linewidth]{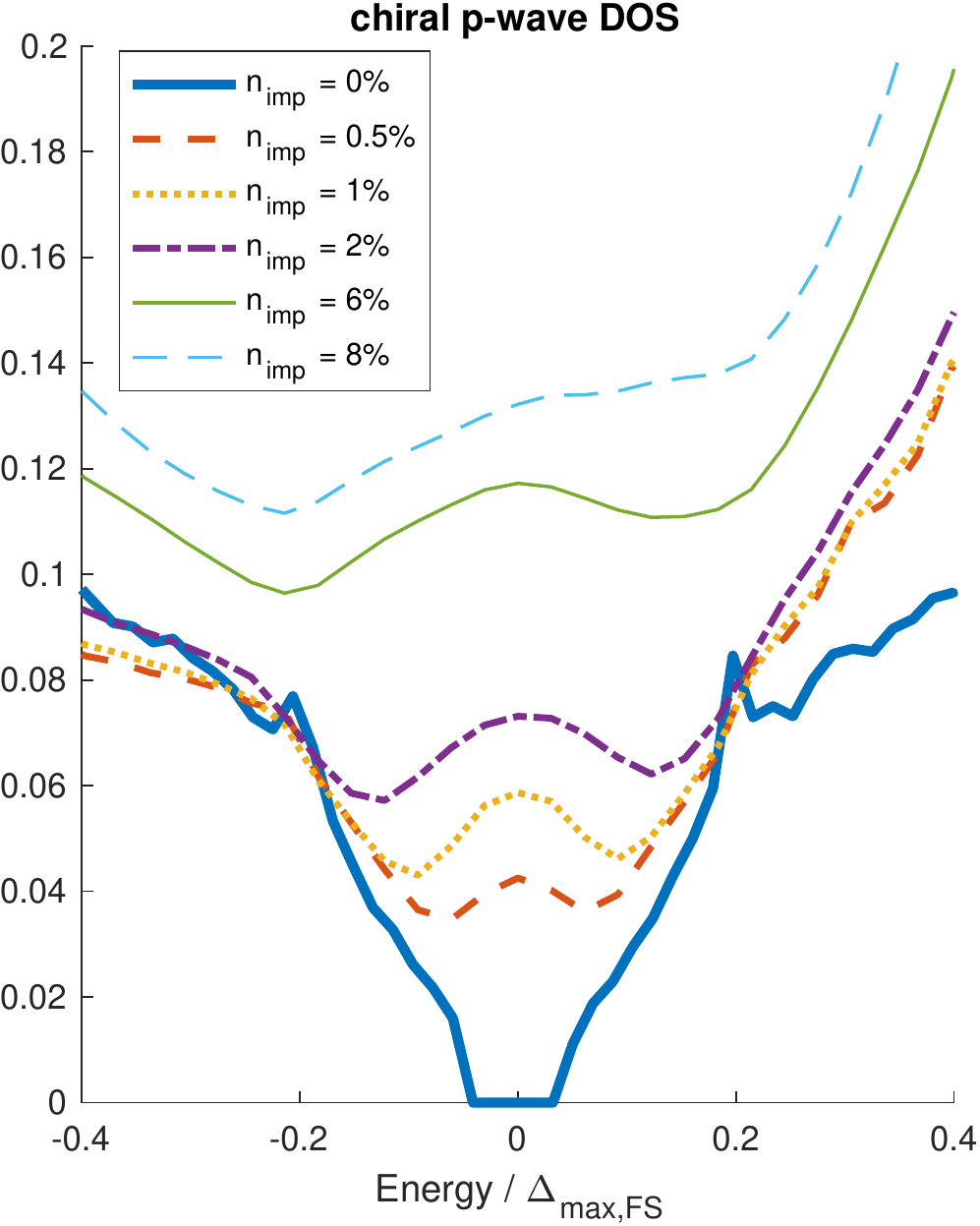}
  \caption{Disorder-averaged DOS for the two-component chiral $p$-wave pairing model
  defined in Sec.~\ref{sec:BdG}. Different colors represent different impurity concentrations $n_{\text{imp}} = 0, 0.5, 1, 2, 6, 8\%$.
  Note that the impurity concentrations here are much smaller than those used for the $s$-wave in Fig.~\ref{fig:DOS_S} because the non-$s$-wave superconductor
  is much more sensitive to disorder.}
  \label{fig:DOS_P}
\end{figure}

In sharp contrast, near-nodes in non-$s$-wave superconductor do not increase, regardless of whether the order parameter is single-component or multi-component.
This can be seen in Fig. \ref{fig:DOS_P}, where the disorder averaged DOS is calculated for
the two-component chiral $p$-wave model with deep gap minima, defined in Sec.~\ref{sec:BdG}.
We see that the addition of a small amount of unitary scattering
results in a filling in of the DOS at zero energy, similar to the single-component $d$-wave case~\cite{Borkowski1994}.
Similar results for a single band chiral $p$-wave superconductor with near-nodes have been obtained in Ref.~\onlinecite{Miyake1999} within the SCTA.
As explained in the introduction, the quite different disorder effects on near-nodes in non-$s$-wave superconductors
come from the fact that the anomalous impurity self energy vanishes.

\section{Thermal Conductivity in BdG} \label{app:kappa}

The thermal conductivity tensor can be computed by the Kubo formula~\cite{Nomura2005, Durst2000} as
\begin{equation}
\kappa_{\mu \nu} = \frac{1}{T} \ \lim_{\Omega \rightarrow 0} \frac{1}{\Omega} \ \lim_{\vec{q} \rightarrow 0} \ \text{Im}
\ \Lambda_{\mu \nu}(\vec{q}, i \omega_m \rightarrow \Omega + i \delta).
\end{equation}
Here $\Lambda_{\mu \nu}(\textbf{q}, i\omega_m)$ is the thermal current-current correlation function at Matsubara frequency
$\omega_m = 2 \pi m / \beta$ with $m \in \mathbb{Z}$ and $\beta=1/(k_B T)$. It is given by
\begin{equation}
\Lambda_{\mu \nu}(\textbf{q}, i\omega_m) =  \int_0^\beta d \tau \ e^{i \omega_m \tau}
\Big\langle T_{\tau} \big[ J_\mu(\textbf{q},\tau) J_\nu(-\textbf{q},0) \big] \Big\rangle,
\end{equation}
where $\langle \cdots \rangle$ denotes thermal ensemble average for a given impurity configuration. $J_\mu(\textbf{q},\tau)$ is
the imaginary-time thermal current operator along the $\mu$ direction and can be approximately split into two parts
\begin{equation}
J_{\mu}(\vec{q},\tau) \approx J^{a}_{\mu}(\vec{q},\tau) + J^{b}_{\mu}(\vec{q},\tau),
\end{equation}
where the superscripts $a$ and $b$ stand for the $d_{xz}$ and $d_{yz}$ orbital contribution, respectively.
In this approximation the current due to the hybridization between $d_{xz}$ and $d_{yz}$ orbitals (from the hopping parameters
$t_{i,i\pm \hat{x} \pm \hat{y}}^{ab}$ and $t_{i,i\pm \hat{x} \mp \hat{y}}^{ab}$ in Eq.~\eqref{eq:H2bandReal} of the main text) has been neglected.
This is a good approximation given that the hybridization is one order of magnitude smaller than the nearest neighbor hopping. For the same reason
we can approximate $\Lambda_{\mu \nu}(\textbf{q}, i\omega_m)$ by
\begin{gather}
\Lambda_{\mu \nu}(\textbf{q}, i\omega_m)\approx \Lambda^{aa}_{\mu \nu}(\textbf{q}, i\omega_m) +\Lambda^{bb}_{\mu \nu}(\textbf{q}, i\omega_m)
\end{gather}
and ignore the cross correlation between current operators of different orbitals.

For the tight-binding model of Eq.~\eqref{eq:H2bandReal}, the thermal current operator of the orbital $a$ is given by~\cite{Paul2003}
\begin{align} \label{eq:currentJmu}
& J^a_\mu(\textbf{q}=0,\tau)  \approx \frac{1}{N_x N_y} \sideset{}{'}\sum_{ij,\sigma} \; (\vec{r}_i -\vec{r}_j)_{\mu} \; t_{ij}^{aa}
\nonumber \\
& \times \frac{1}{2} \bigg[ \frac{\partial c^\dagger_{i a \sigma}(\tau)}{\partial \tau} c_{j a \sigma}(\tau) - c^\dagger_{i a \sigma}(\tau) \frac{\partial c_{j a \sigma}(\tau)}{\partial \tau} \bigg],
\end{align}
where $c_{ia\sigma}^\dagger(\tau)=e^{\tau H} c_{ia\sigma}^\dagger e^{-\tau H}$ and  $c_{ia\sigma}(\tau)=e^{\tau H} c_{ia\sigma} e^{-\tau H}$ with $H$ the mean
field BdG Hamilton. The prime in the summation means only sites $i$ and $j$ that are connected by nonzero hoppings $t_{ij}^{aa}$ are summed over. In this formula,
$(\vec{r}_i -\vec{r}_j)_{\mu} \; t_{ij}^{aa}$ is the electron hopping velocity operator along the $\mu$-direction, where $\vec{r}_i$ is the
coordinate of site $i$ on the square lattice. Also, Eq.~\eqref{eq:currentJmu} only includes the kinetic energy part contribution to the thermal current,
as indicated by the $\approx$ sign, and neglects the potential energy part~\cite{Paul2003}, or the superconducting order parameter part~\cite{Durst2000}, which is appropriate for
$\mathrm{Sr_2 Ru O_4}$, since its superconducting transition temperature $T_c$ is much smaller than the band hoppings.

Substituting Eq.~\eqref{eq:currentJmu} into the definition of $\Lambda_{\mu \nu}^{aa}$ leads to
\begin{widetext}
\begin{align}
& \Lambda_{\mu \nu}^{aa}(\vec{q} =0,i \omega_m) = \frac{1}{4 (N_x N_y)^2} \sideset{}{'}\sum_{ij,\sigma} \sideset{}{'}\sum_{i^\prime j^\prime,\sigma'}
\big[ (\vec{r}_i -\vec{r}_j)_{\mu} \;  t_{ij}^{aa} \big] \big[ (\vec{r}_{i^\prime} -\vec{r}_{j^\prime})_{\nu} \;  t_{i^\prime j^\prime}^{aa} \big] \nonumber \\
& \times \int_0^\beta d \tau \ e^{i \omega_m \tau}
\bigg \langle T_\tau  \bigg\{ \dot{c}^\dagger_{i a \sigma} c_{j a \sigma}    \;  \dot{c}^\dagger_{i^\prime a \sigma'} c_{j^\prime a \sigma'}
                              -  \dot{c}^\dagger_{i a \sigma} c_{j a \sigma}  \;  c^\dagger_{i^\prime a \sigma^\prime} \dot{c}_{j^\prime a \sigma'}
                              -  c^\dagger_{i a \sigma} \dot{c}_{j a \sigma}   \; \dot{c}^\dagger_{i^\prime a \sigma'} c_{j^\prime a \sigma'}
                              +  c^\dagger_{i a \sigma} \dot{c}_{ j a \sigma}   \; c^\dagger_{i^\prime a \sigma'}\dot{c}_{j^\prime a \sigma'}
\bigg\}  \bigg\rangle, \label{eq:cccc}
\end{align}
\end{widetext}
where
\begin{align} \label{eq:Heisenberg}
\dot{c}_{i a \sigma} & \equiv \frac{\partial c_{i a \sigma}}{\partial \tau} = \big[H, c_{i a \sigma} \big]  \nonumber \\
&  = \sum_n  E_n \big[v^{*}_{n a}(i) \gamma^\dagger_{n \overline{\sigma}} -u_{n a}(i) \gamma_{n \sigma}\big].
\end{align}
In Eq.~\eqref{eq:cccc} we have suppressed the $\tau$ dependence of each operator product for brevity. Eq.~\eqref{eq:Heisenberg} is obtained
by using the Bogoliubov transformation from Eq.~\eqref{eq:Bogoliubov} and the diagonalized BdG Hamiltonian
$H = \sum_{n,\sigma} E_n \gamma^\dagger_{n \sigma} \gamma_{n \sigma}$.

Next we plug Eq.~\eqref{eq:Bogoliubov} and \eqref{eq:Heisenberg} into Eq.~\eqref{eq:cccc}, carry out the expectation value of each
term in Eq.~\eqref{eq:cccc} using Wick's theorem, complete the imaginary time integral, and then perform the analytic continuation
$i\omega_m \rightarrow \Omega+ i\delta$. The final result of $\Lambda_{\mu \nu}(\vec{q} =0,\Omega+ i\delta)$ is fully in terms of the eigenvalues, $E_n$, and eigen-functions, $(u_{n a \sigma}, v_{n a \sigma})$, of the BdG Hamiltonian $H$
so that it can be evaluated numerically. Although the derivation is quite straightforward,
the final expression is quite lengthy so we do not present it here.

\end{appendix}


%

\end{document}